\documentclass[12pt,epsf,epsfig,psfig]{article}
\usepackage{graphicx}
\usepackage{epsfig}
\usepackage{cite}
\def\J{$J/\psi$}
\def\j{J/\psi}

\def\x{\chi}
\def\P{$\psi'$}
\def\p{\psi'}
\def\U{$\Upsilon$}

\def\C{c{\bar c}}

\def\P{$\psi'$}
\def\p{\psi'}
\def\U{$\Upsilon$}

\def\C{c{\bar c}}

\def\e{\epsilon}

\def\NP{{ Nucl.\ Phys.\ }}
\def\PL{{ Phys.\ Lett.\ }}
\def\PR{{ Phys.\ Rev.\ }}

\def\PRL{{ Phys.\ Rev.\ Lett.\ }}

\def\ZP{{ Z.\ Phys.\ }}
\def\EPJ{{Eur.\ Phys.\ J.\ }}

\def\e{\epsilon}

\def\NP{{ Nucl.\ Phys.\ }}
\def\PL{{ Phys.\ Lett.\ }}
\def\PR{{ Phys.\ Rev.\ }}

\def\PRL{{ Phys.\ Rev.\ Lett.\ }}

\def\ZP{{ Z.\ Phys.\ }}
\def\EPJ{{Eur.\ Phys.\ J.\ }}

\def\be{\begin{equation}}
\def\ee{\end{equation}}
\def\lsim{\raise0.3ex\hbox{$<$\kern-0.75em\raise-1.1ex\hbox{$\sim$}}}
\def\gsim{\raise0.3ex\hbox{$>$\kern-0.75em\raise-1.1ex\hbox{$\sim$}}}

\begin{document}

\parindent=0pt 

{\small ~
 \hfill BI-TP 2013/22}
~~

\vskip1cm

\centerline{\Large \bf Probing the States of Matter in QCD$^*$}

\vskip1cm

\centerline{\bf Helmut Satz}

\bigskip

\centerline{Fakult\"at f\"ur Physik, Universit\"at Bielefeld}

\centerline{D-33501 Bielefeld, Germany}



\vskip1cm

\centerline{\bf \large Abstract:}

\bigskip

The ultimate aim of high energy heavy ion collisions is to study  
quark deconfinement and the quark-gluon plasma predicted by quantum
chromodynamics. This requires the identification of observables calculable
in QCD and measurable in heavy ion collisions. I concentrate on three
such phenomena, related to specific features of strongly interacting matter.
The observed pattern of hadrosynthesis corresponds to that of an ideal 
resonance gas in equilibrium at the pseudo-critical temperature determined 
in QCD. The critical behavior of QCD is encoded in the fluctuation patterns 
of conserved 
quantum numbers, which are presently being measured. The temperature of the 
quark-gluon plasma can be determined by the dissociation patterns of the 
different quarkonium states, now under study at the LHC for both charmonia
and bottomonia.







\vskip1cm

\section{\large Feynman's Broken Watch}

\medskip

The high energy heavy ion program was initiated in the nineteen-eighties
at CERN and at Brookhaven, and it had a well-defined aim: to produce and 
study in the la\-bo\-ratory the deconfined state of matter, the quark-gluon 
plasma, predicted some years earlier by quantum chromodynamics. The charge 
was thus to create a new state of matter through high energy collisions of 
heavy nuclei. 

\medskip

The investigation of such collisions is a very multi-faceted enterprise.
It involves initial state aspects, parton structure and its limit as
color glass, multiple parton interactions, non-equilibrium evolution, 
thermalization questions, hydrodynamic expansion, viscosity and flow, and 
much more. The final creation of deconfined thermal systems through such 
interactions remains a rather speculative issue, and the view of sceptics 
was perhaps best summarized by Richard Feynman when he said: ``if I throw my 
watch against the wall, I get a broken watch, not a new state of matter''. 
The problem has two inherent aspects. On the one hand, we have to show that
the collision leads to a medium with large scale collective behavior, 
something one would call matter; on the other hand, we want the initial 
state of this thermal system to be the QCD plasma of quarks and gluons.

\medskip

---------------------------------------------------------------------
\medskip

* Survey talk given at the {\sl 26$^{\rm th}$ International Symosium
on Lepton Photon Interactions at High Energies}, San Francisco,
California, USA, June 24 - 29, 2013.

\newpage

The mentioned non-equilibrium issues are difficult, if not impossible, to 
account for in terms of first principle QCD calculations. Experimentally, 
the observation of hydrodynamic flow, both radial and elliptic, shows
the presence of collective effects. Moreover, the resulting medium is 
extremely dense, leading to a strong quenching of high transverse momentum 
jets. Much interesting theoretical work has been carried out on the
interpretation of these results, and this has led to something one might 
call a split of paradigms. On one hand, one can attempt to model the 
dynamical collision 
process in its various stages from partonic beginning to hadronic end 
and then check to see if the data agree with the model. As interesting
as such an approach is, it does not really assure us that the produced 
medium is indeed the quark-gluon plasma described in non-perturbative 
QCD studies. To conclude that, there seems to be only one way: to calculate 
in equilibrium statistical QCD some observable features and then test if 
these features indeed arise in high energy nuclear collisions. This 
approach, if successful, will tell us whether we have fully 
carried out the charge assigned to us at the start of the heavy ion
program. The aim of my survey will be this line of study, to identify
{\sl ab initio} results from statistical QCD which can be compared to data 
from high energy heavy ion collisions. Where are we at present in this 
specific task?

\medskip

I will in particular cover three issues: are there experimental indications 
for thermal equilibrium of the produced hadronic medium, and if so, how can 
we look in heavy ion collisions for the critical features of QCD at the 
quark-hadron transition? Finally, on the deconfined side, how can we measure 
in a collision environment the temperature of the produced deconfined 
environment, the expected hot quark-gluon plasma? The first two of these 
topics are presently being addressed in fluctuation studies at the CERN-LHC,
in a dedicated experiment at the CERN-SPS, and by the beam energy scan at 
RHIC. The last is a central theme for the heavy ion program of three LHC 
and two RHIC experiments. With a consideration of these topics, my 
heavy ion theory report will moreover be quite complementary to those 
given at the last two Lepton-Photon conferences \cite{Renk,Olli}, which 
have concentrated more on the dynamical evolution aspects of the produced 
medium.
  
\section{\large Hadrosynthesis and Freeze-Out}

\begin{figure}[htb]
\centerline{\epsfig{file=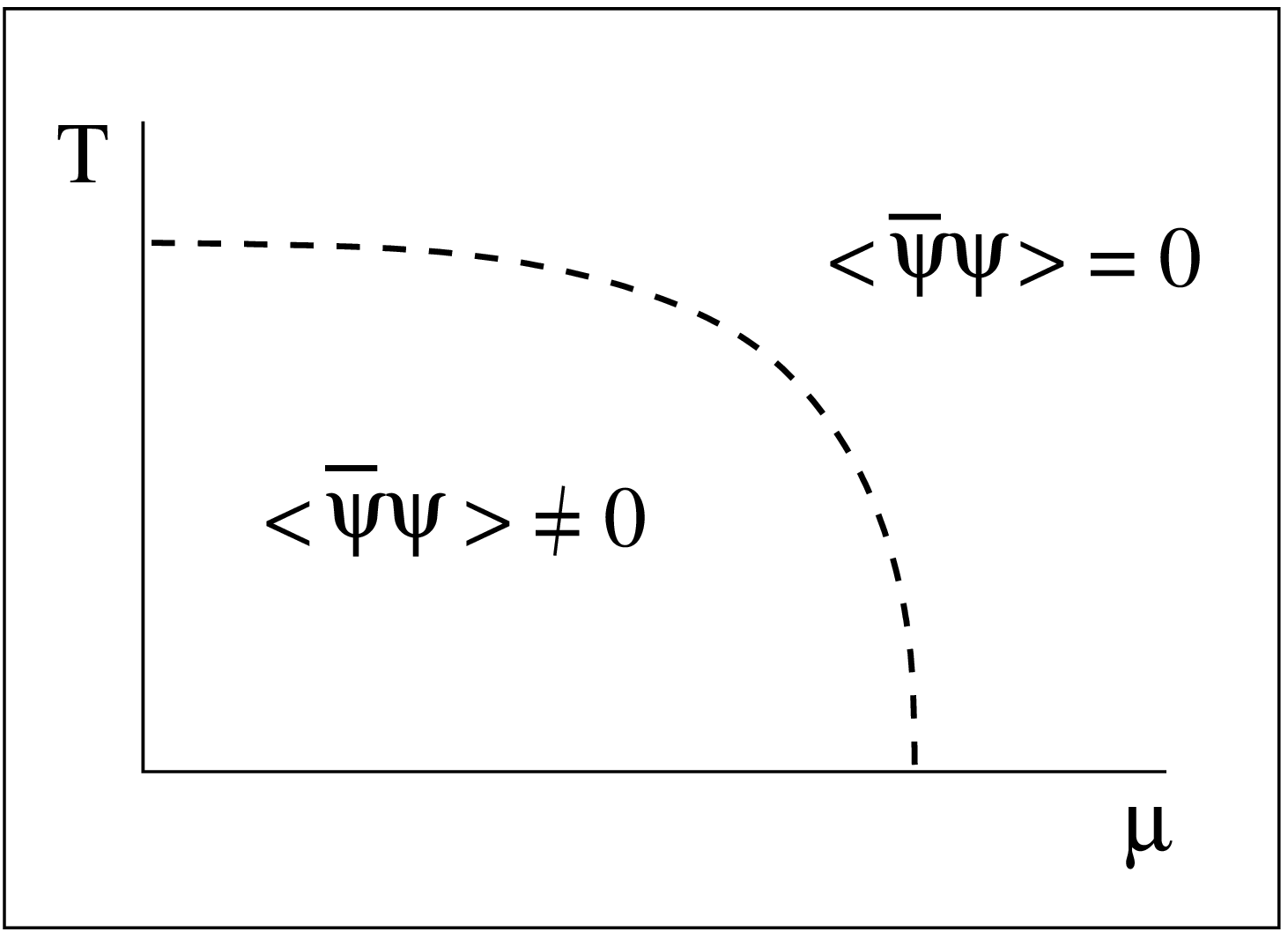,width=4.7cm}\hskip1.5cm
\epsfig{file=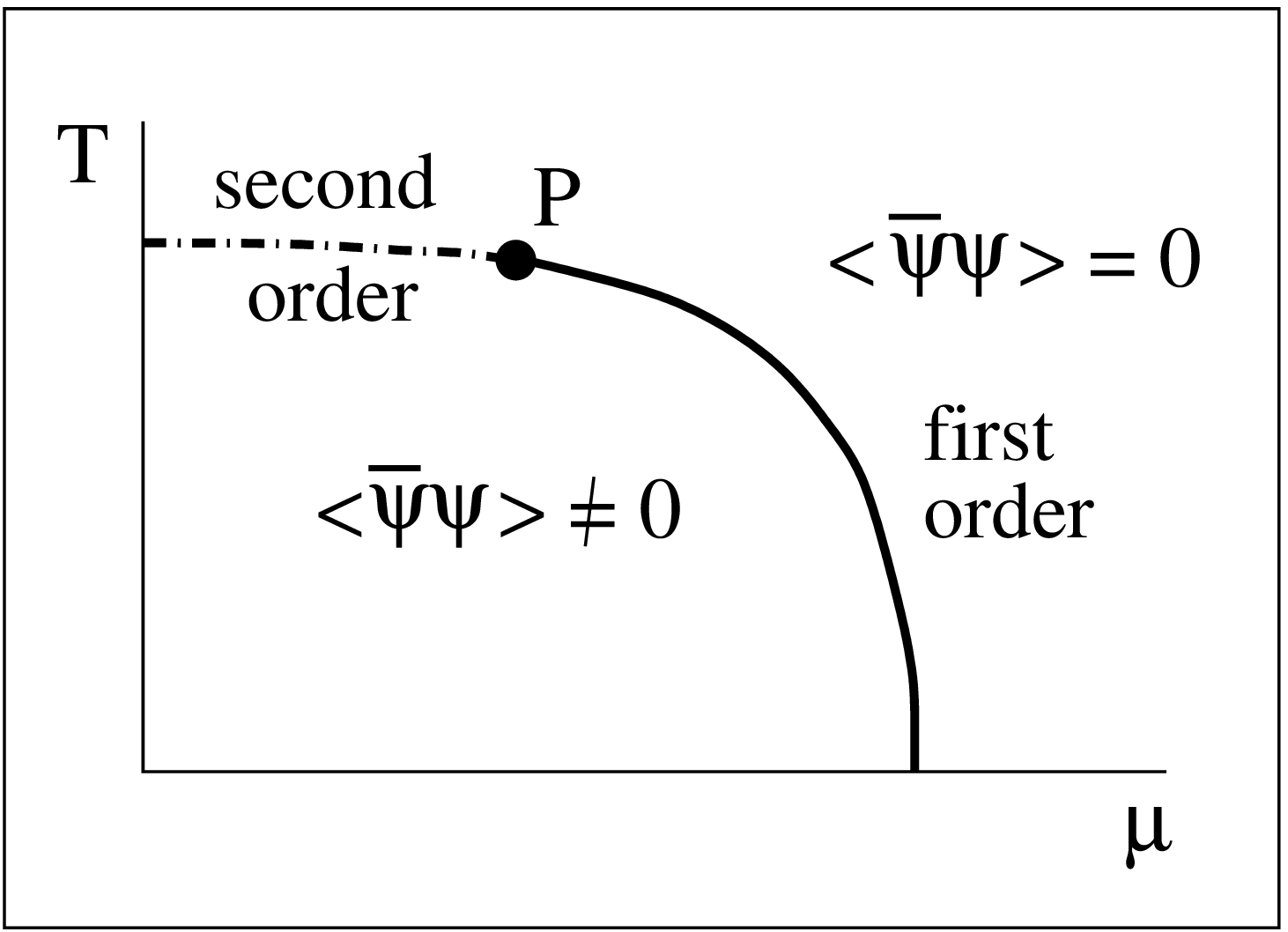,width=4.7cm}}
\vspace*{0.2cm}

\hskip4.6cm (a) \hskip5.6cm (b)

\vspace*{0.2cm}
\caption{(a) Generic phase diagram for QCD of two massless quark 
flavors; (b) specific transition structure: second order (dashed line),
tricritical point (P), first order (solid line).}

\label{phase0}
\end{figure}

For two massless quark flavors, strongly interacting matter
as function of temperature $T$ and baryochemical potential $\mu$ shows
a two-phase structure, defined by chiral symmetry. At low $T$ and $\mu$,
the expectation value of the chiral condensate, $\chi(T,\mu) = \langle 
{\bar \psi} \psi \rangle$, is non-zero, the chiral symmetry of the QCD 
Lagrangian is spontaneously broken; with increasing $T$ and/or $\mu$ we 
reach a critical line in the $T-\mu$ plane at which chiral symmetry is 
restored. The resulting generic phase diagram is shown in Fig.\ 
\ref{phase0}(a). More specifically, the 
transition at $\mu=0$ is conjectured to be of second order and in the 
universality class of the three-dimensional O(4) spin system \cite{PW};
lattice studies support this and moreover have the second order behavior
continue up to some (small) values of $\mu$ \cite{latticePW2}. 
At low temperatures, various 
arguments suggest that the transition is of first order \cite{firstorder}. 
We thus expect a tricritical point $P$ at the position in the $T-\mu$ plane 
where the two transition lines meet \cite{tricritical,latticePW1}. 
The corresponding phase diagram is shown in Fig.\ 
\ref{phase0}(b). At small $\mu$, the transition line not only defines chiral 
symmetry breaking and restoration, but it also separates the state of 
deconfined quarks and gluons from an interacting hadronic medium. 

\medskip

\medskip

\begin{figure}[htb]
\centerline{\epsfig{file=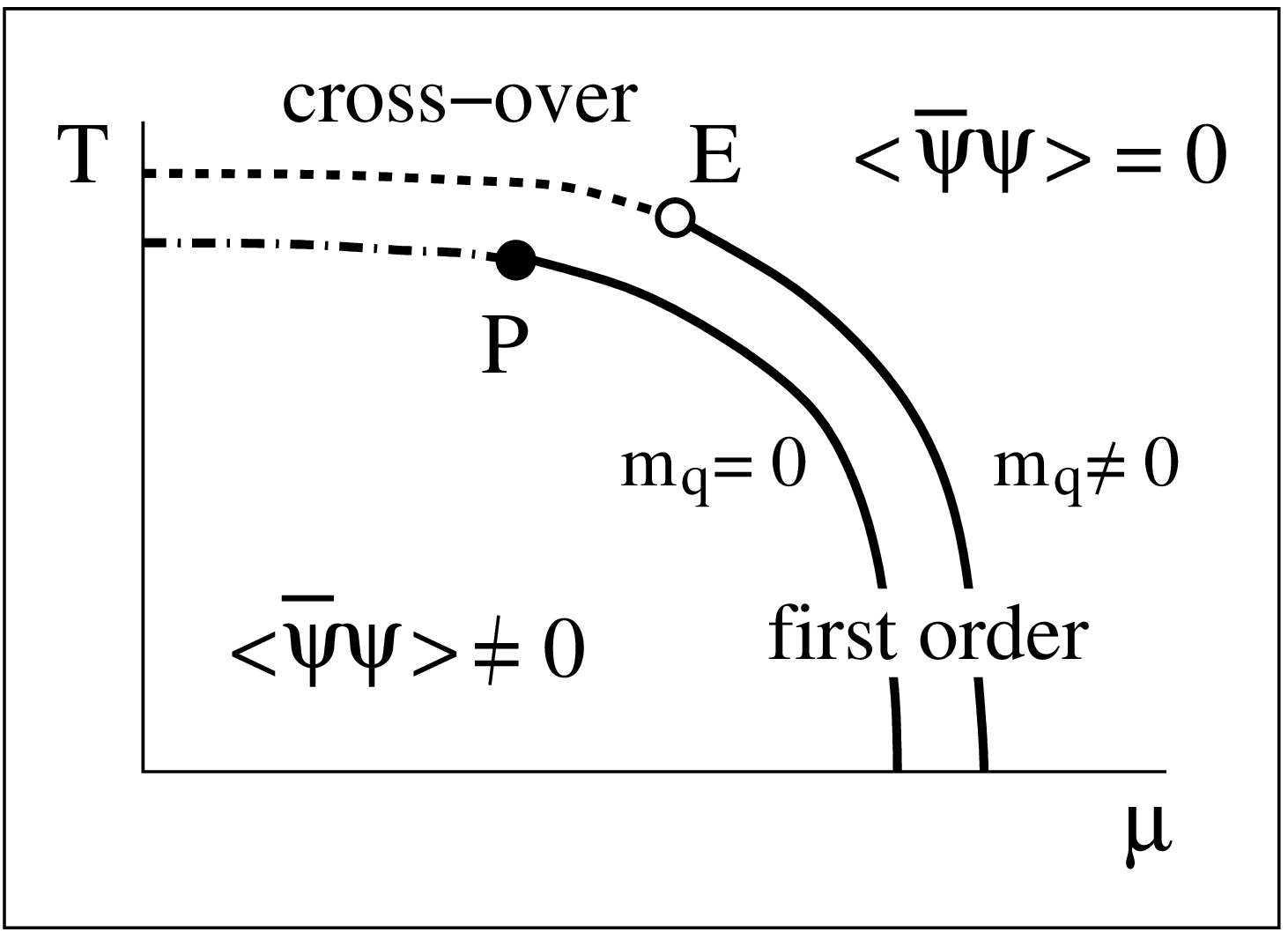,width=4.7cm}\hskip1.5cm
\epsfig{file=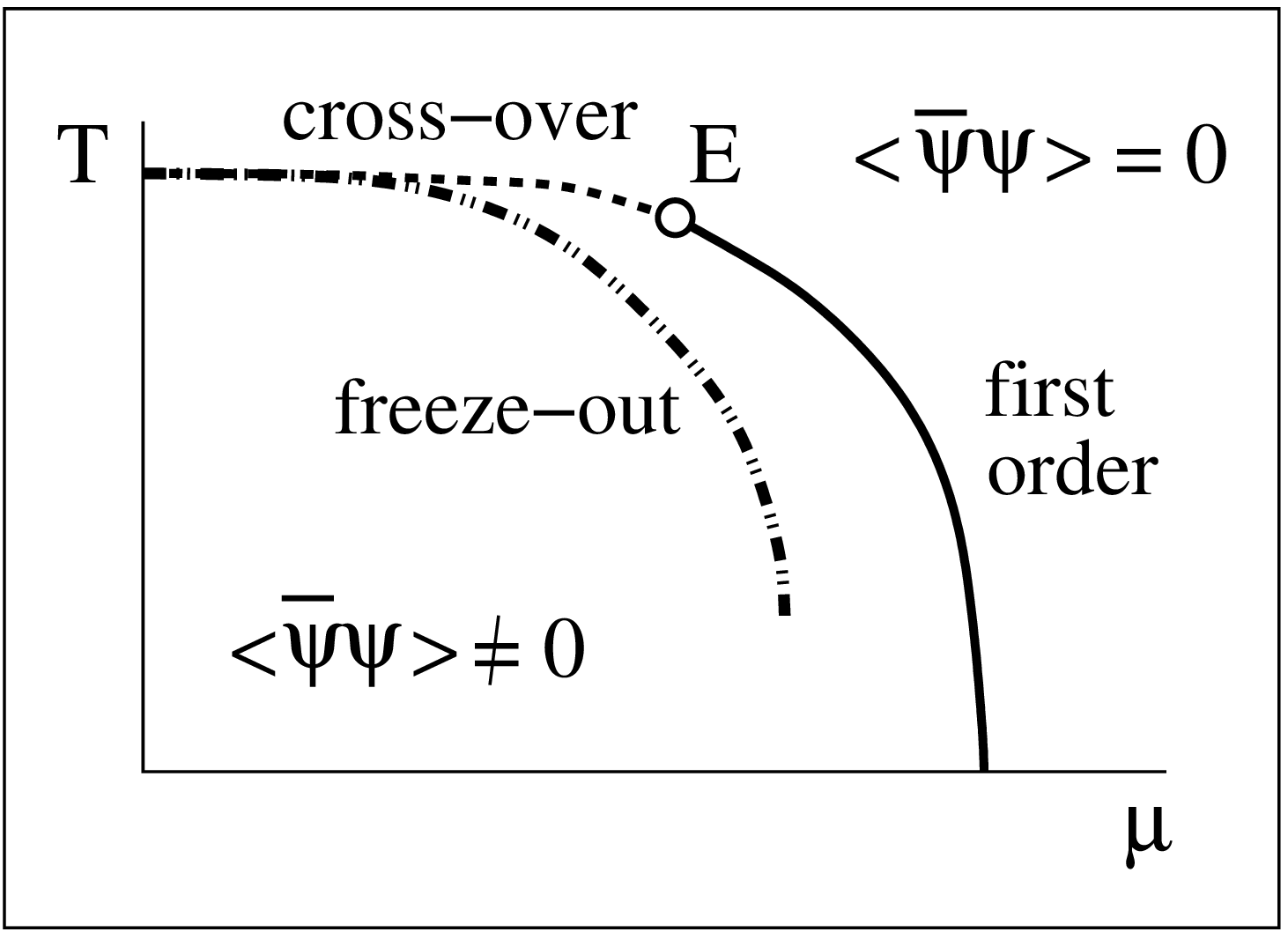,width=4.7cm}}
\vspace*{0.2cm}

\hskip4.6cm (a) \hskip5.6cm (b)

\caption{(a) Phase diagram for QCD with $m_q\not=0$ compared to the case
$m_q=0$; (b) expected freeze-out line compared to the transition line for
$m_q\not=0$.}
\label{phase1}
\end{figure}

The real world has $u$ and $d$ quarks of small but finite masses $m_q$,
and the presence of these masses affects continuous critical behavior in 
much the same way as an external field does to a spin system - it turns 
singular behavior into a pseudo-critical rapidly varying cross-over of
the relevant physical variables. On the other hand, a first order transition 
can be modified parametrically, but not removed completely, so that such a 
transition remains present also for small but finite quark masses $m_q$. 
As a result we obtain the pattern shown in Fig.\ 
\ref{phase1}(a); the endpoint $E$ of the first order line is now simply 
critical, with the $Z_2$ exponents of the three-dimensional Ising model, 
at which the pseudo-critical cross-over starts. 

\medskip

\begin{figure}[htb]
\centerline{\epsfig{file=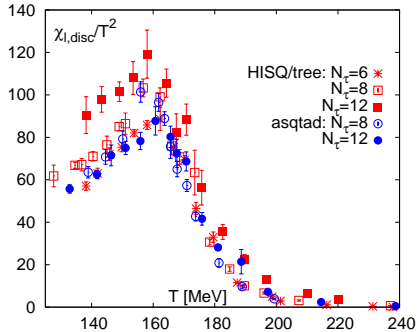,width=6cm}}
\vspace*{0.2cm}

\caption{Temperature variation of the 
chiral susceptibility in (2+1) QCD with physical quark 
masses \cite{temp}.}
\label{chiral-disc}
\end{figure}

Adding a third heavier $s$ quark does not significantly change the pattern, 
so the behavior shown in Fig.\ \ref{phase1}(a) holds for this case as well. 
For a system with 2+1 physical quark mass values, the temperature of the 
transition at $\mu=0$ is in the latest lattice studies \cite{temp} established
by calculating the continuum limit of the chiral susceptibility, i.e., the 
derivative of the chiral condensate with respect to the light quark mass for 
$m_q \to 0$. It is found to peak at $T_H = 154 \pm 9$ MeV, see Fig.\ 
\ref{chiral-disc}. 

\medskip

If the interactions in the hadronic system appearing just below the transition
are resonance-dominated, the newly formed confined medium can be represented 
as an ideal gas of all possible resonance states \cite{BU,DMB} at one common 
``freeze-out'' temperature $T_f$. The relative abundances of the different
hadron species are then determined at this point; subsequent cooling
can change the abundances only through decay, not through interaction. In 
the low baryon density limit, i.e., at small or vanishing $\mu$, resonance 
dominance is a good assumption; we recall the success of the dual resonance 
model. With increasing $\mu$, however, baryon-baryon interactions begin to 
play a role, and these are not of resonance nature; hence the freeze-out 
curve is expected to fall eventually below the chiral transition curve, 
as shown in Fig.\ \ref{phase1}(b).

\medskip

\begin{figure}[htb]
\centerline{\epsfig{file=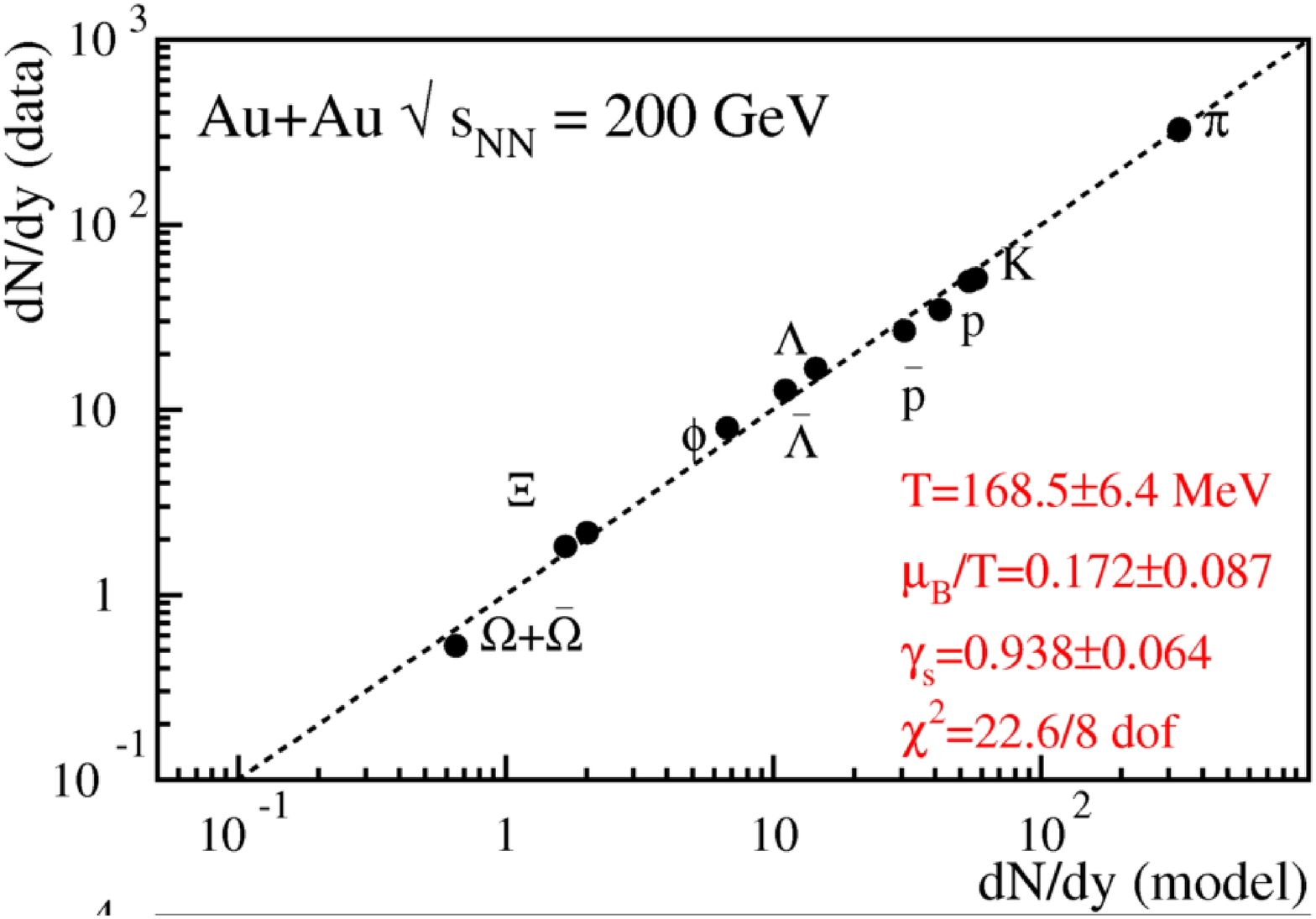,width=7.3cm}\hskip1cm
\epsfig{file=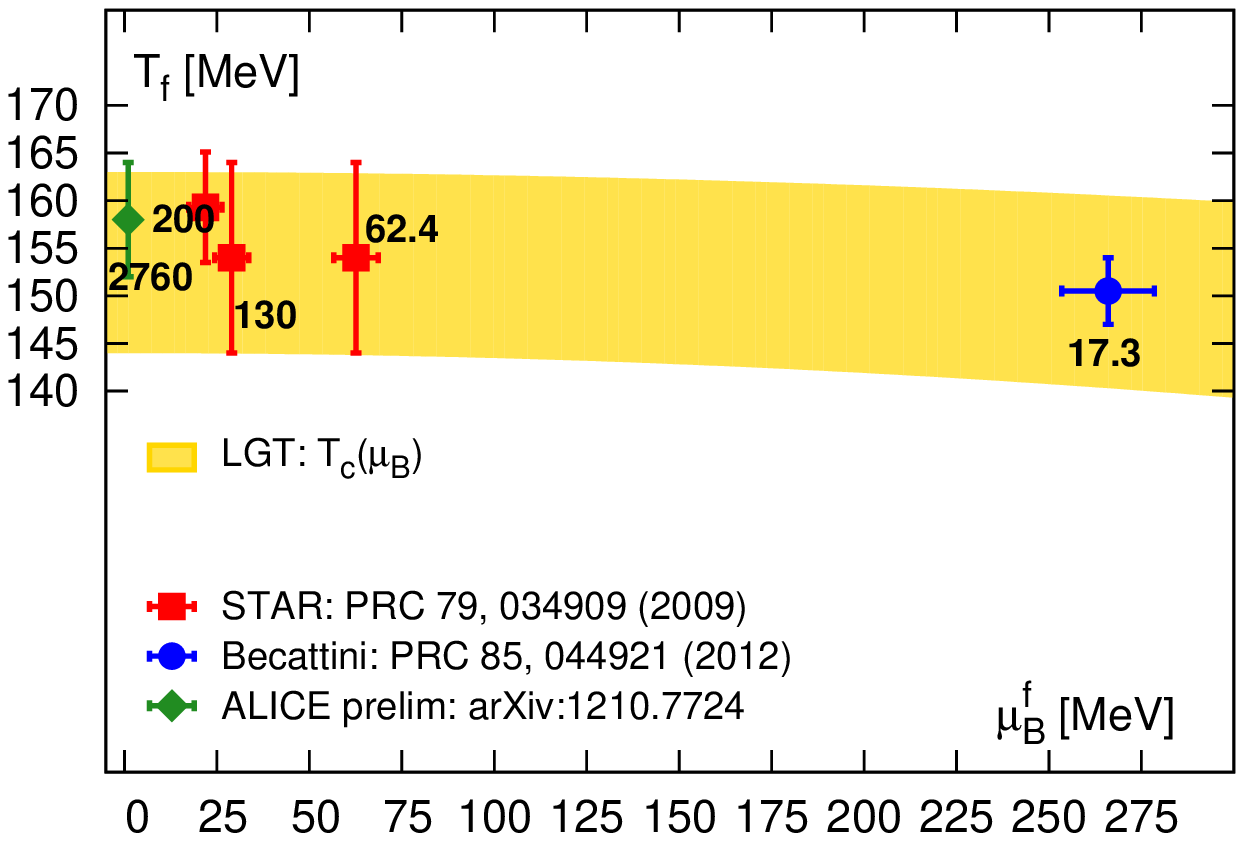,width=6.3cm}}
\hskip3.8cm (a) \hskip7.8cm (b)
\caption{(a) Species abundances in Au-Au collisions; (b) freeze-out parameters
in heavy ion collisions at different energies vs. lattice QCD results.}  
\label{Becaplot}
\end{figure}

In an ideal gas of hadronic resonances at fixed freeze-out temperature $T_f$, 
the abundances of the various species are specified by the corresponding
phase space weights. The relative abundance of states $i$ and $j$ is
thus given by
\be
{N_i \over N_j} = \left({d_i \over d_j}\right)
 \left({m_i  \over m_j}\right)^2 {K_2(m_i/T_f) \over 
K_2(m_j/T_f)} \simeq \left({d_i \over d_j}\right)
\left( {m_i \over m_j}\right)^{3/2}\!
\exp\{-(m_i - m_j)/T_f\},
\label{1.1}
\ee
where $m_i$ denotes the mass and $d_i$ specifies the intrinsic degeneracy 
(charge, spin) of the state. We have here assumed non-strange mesons; 
baryon number and strangeness lead to additional factors, to which we
shall return shortly. The striking feature observed in high energy heavy 
ion collisions is that the relative abundances of all observed hadrons, 
more than a dozen species, are well-described in terms of such a resonance 
gas, with one common freeze-out temperature $T_f = 160 \pm 10$ MeV
\cite{11AA1,11AA2,11AA3,11AA4,11AA5,11AA6}. In Fig.\ \ref{Becaplot} we 
give an illustration of the agreement. The slight temperature differences
between the two plots arise from an introduction of non-resonant
baryon interactions in a more recent version of the model \cite{stock}.
The distribution of the pieces of 
the broken watch is thus given by an ideal gas of fixed tempe\-ra\-ture: 
the collision does produce something like hadronic matter, and the
hadrosynthesis occurs at just the transition temperature predicted 
by QCD.

\medskip

There is an interesting caveat to be added here. The abundances of the 
hadron species
produced in elementary interactions ($pp$, $p\bar p$, $e^+e^-$) are also
reproduced with the same or better precision by a thermal resonance gas
of the same temperature of 160 MeV, although here one presumably does not 
create a genuine thermal medium. In these interactions, however, there is 
one clear deviation from the predictions of an ideal resonance gas model:
the abundances of strange particles are systematically lower than the 
predictions. This has been accounted for phenomenologically by the 
introduction of a strangeness suppression factor $\gamma_s \simeq 0.5 - 
0.6$ \cite{raf}; the production rate of a hadrons containing $n$ strange 
quarks are then reduced by a factor $\gamma_s^n$. This one additional
parameter then allows an excellent account of all elementary hadroproduction.
The origin of this thermal behavior has been an enigma for many years;
it is most likely resolved by local stochastic production, initiated by
quark tunnelling through the confinement event horizon \cite{CKS}. Such
a scheme in fact provides automatically the strangeness suppression
observed for elementary interactions \cite{BCMS}.

\medskip

The mere observation of thermal hadron abundances thus does not establish 
that a truly thermal medium was formed. A first indication that this is in 
fact the case in heavy ion collisions is indicated by the convergence
of the suppression factor $\gamma_s \to 1$, which arises from a statistical
averaging over the production processes due to the different nucleon-nucleon
collisions in the interaction \cite{BS}. All abundances, including those
of strange hadrons, are now indeed given by an ideal resonance gas. 

\medskip

To further corroborate this claim, one has studied the temperature dependence 
of conserved quantum numbers in an ideal hadron gas and compared this to both 
lattice results and heavy ion data. We consider as an illustration the 
baryon number behavior. The pressure of the hadron gas is then 
given by the sum over all resonance species up to some mass of around
2.5 GeV (a further increase does not lead to significant changes),
\be
{P(T,\mu) \over T^4}
= {1\over \pi^2} \sum_i d_i (m_i/T)^2 K_2(m_i/T) \cosh(B_i\mu/T),
\label{1.2}
\ee
where $B_i$ denotes the baryon number of the species and $\mu$ the
corresponding baryochemical potential. The meson contribution, with $B_i=0$, 
depends on $T$ only; in general, however, a fit of the observed species
abundances thus specifies the two freeze-out parameters $T_f$ and $\mu_f$. For 
collision energies $\sqrt s \geq 20$ GeV, the temperature is found to have the 
mentioned value of some 160 - 165 MeV; the baryochemical freeze-out value 
$\mu_f$ decreases from about 220 MeV at top SPS energy (20 GeV) to 25 MeV 
at top RHIC energy (200 GeV) and essentially zero at the LHC.
As expected, a variation of the collision energy thus 
results in a considerable variation of the baryochemical potential; increasing
the collision energy leads to more and more nuclear tranparency and thus
to lower baryon density. This effect
can be used to study fluctuations in baryon number. The generalized 
$n-th$ order baryon number cumulant, 
\be
\chi_B^{(n)}(T,\mu) = {\partial^n (P/T^4) \over \partial(\mu/T)^n},
\label{1.3}
\ee
is readily calculated from eq.\ \ref{1.2}, and keeping in mind that
only baryons with $B=1$ enter, one finds relations of the form
\be
{\chi_B^{(3)} \over \chi_B^{(1)}} =
{\chi_B^{(4)} \over \chi_B^{(2)}} = 
{\chi_B^{(5)} \over \chi_B^{(3)}} = ... = 1
\label{1.4}
\ee
for ratios two units apart. For those separated by one unit, one has
\be
{\chi_B^{(2)} \over \chi_B^{(1)}} = \coth(\mu/T),~
{\chi_B^{(3)} \over \chi_B^{(2)}} = \tanh(\mu/T),~
{\chi_B^{(4)} \over \chi_B^{(3)}} = \coth(\mu/T),
\label{1.5}
\ee
and so on. 
If the confinement transition has led to a hadronic resonance gas in
equilibrium, all memories of previous stages are lost, and so an
agreement with relations (\ref{1.4})/(\ref{1.5}) would indeed confirm the 
production of {\sl hadronic matter} of a freeze-out temperature
$T_f = T_H$ equal to the transition value.

\medskip

\begin{figure}[htb]
\centerline{\epsfig{file=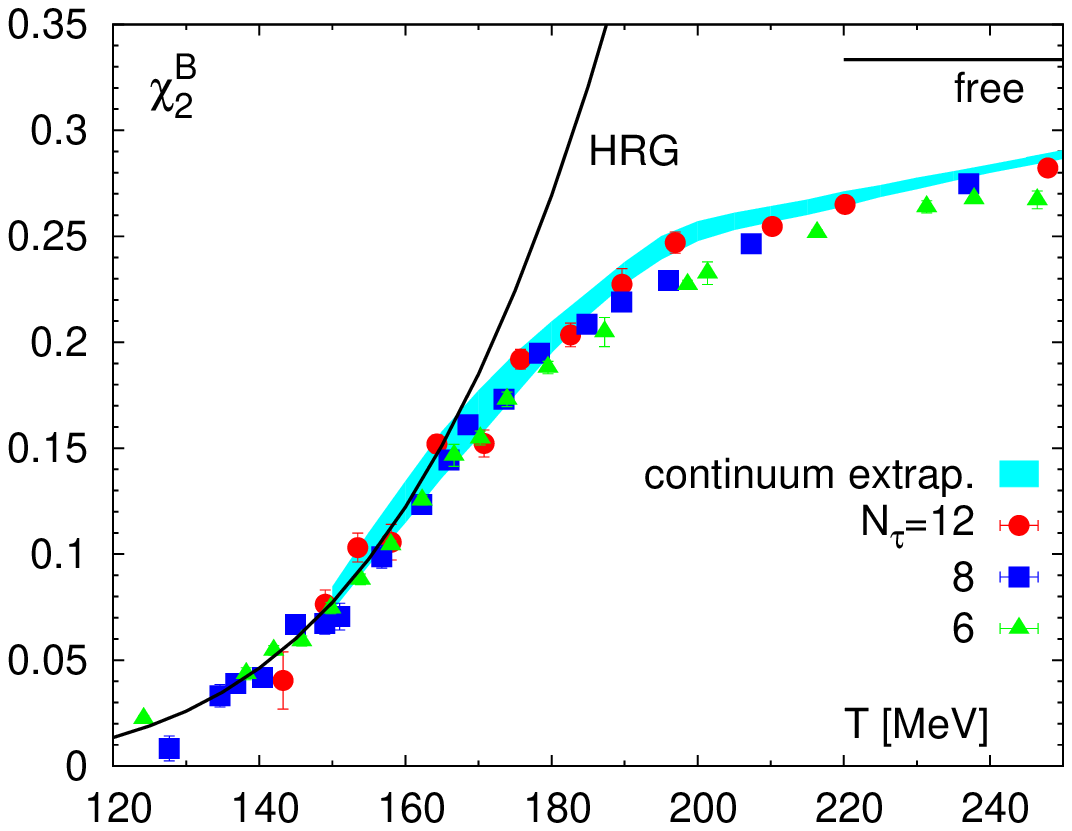,width=5.4cm}\hskip1cm
\epsfig{file=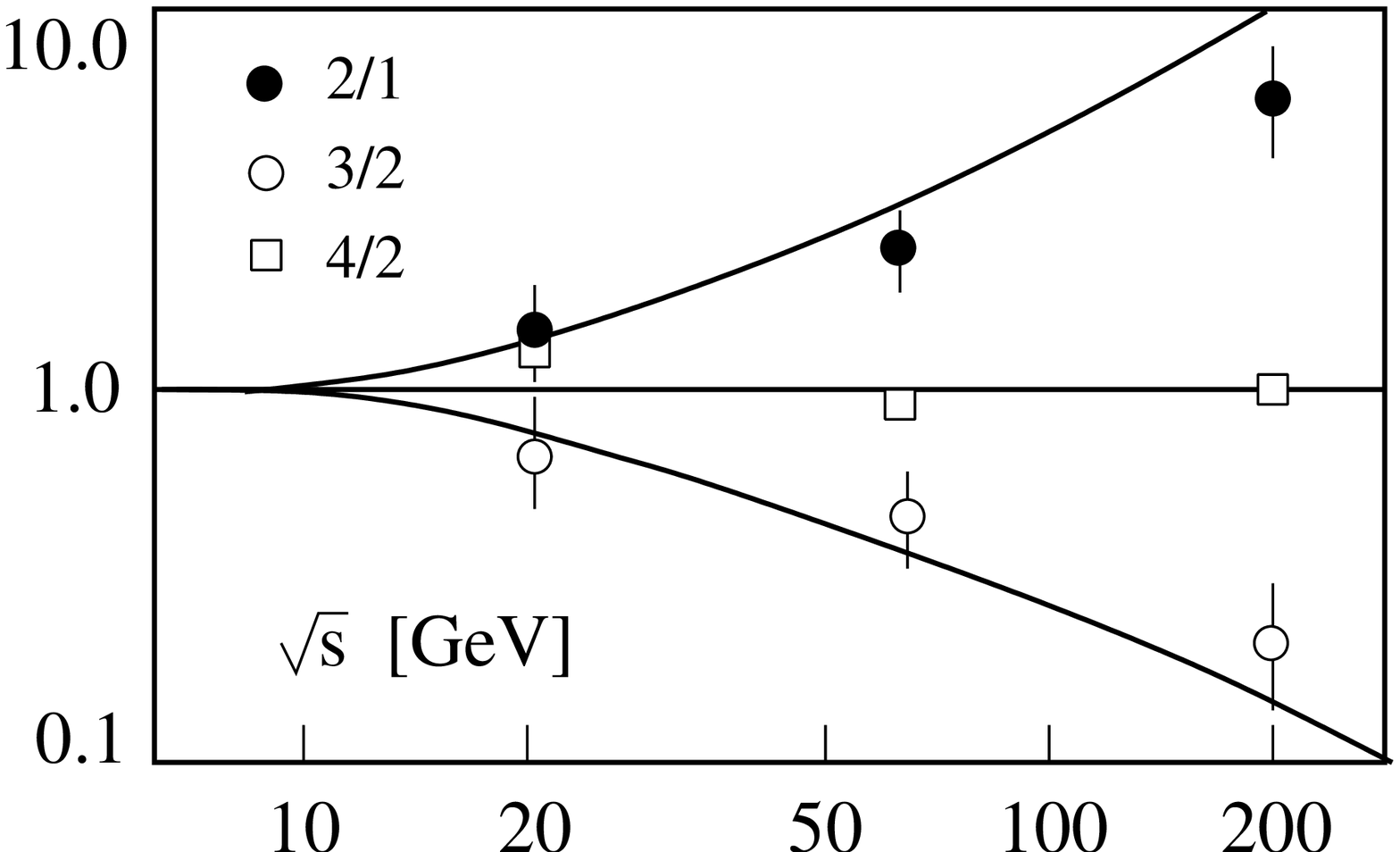,width=6cm}}
\vspace*{0.2cm}

\hskip4.5cm (a) \hskip6cm (b)

\caption{(a) Theory: lattice results (2+1 quark flavors, $\mu=0$)
for the second order baryon number 
cumulant, compared to the hadron resonance gas form.
(b) Experiment: STAR data \cite{star} on baryon cumulant ratios, eqns.\
(\ref{1.4}) and (\ref{1.5}), as function of collision energy $\sqrt s$,
vs.\ hadron resonance gas predictions \cite{HRG} (solid lines).} 
\label{HRG}
\end{figure}

Lattice studies have so far shown that such an ideal resonance gas behavior 
is also what QCD predicts. As an illustration, we show in Fig.\ \ref{HRG}(a)
results for the second baryon number cumulant $\chi^B_{2}$ in (2+1) flavor 
QCD, compared to the corresponding hadron resonance gas prediction. Up to the 
transition temperature, the agreement is excellent. Experimentally,
the cumulants can be measured in terms of higher moments of the 
baryon number dispersion, $N_B - \langle N_B \rangle$, and such data has
been obtained by the STAR collaboration at RHIC \cite{HRG}. Species 
abundances determine $T_f$ and $\mu(T_f)$ at collision energies of 
19.6, 62.4 and 200 GeV.
The baryon number fluctuations allow a determination of the cumulant
ratios, and the results for three ratios are compared in Fig.\ \ref{HRG}(b)
to the resonance gas predictions (\ref{1.4}) and (\ref{1.5}) \cite{HRG}. 
The agreement is also seen to be quite good.

\medskip

We thus find that high energy heavy ion collisions produce a medium which 
can be considered as hadronic matter in equilibrium, formed at 
the pseudo-critical hadronization temperature predicted by lattice QCD.
And cumulant studies show that also correlations in this medium appear to
follow the pattern of an ideal resonance gas, which for the lower orders
is again in accord with QCD. But evidently we would like more: we want to 
find in the collision data some sign of the actual transition, of something 
like critical behavior. Sufficiently close to a continuous transition,
correlations appear at all scales, the correlation length diverges,
and in QCD this must produce strong deviations from the ideal hadron gas 
behavior just studied. What is ``close enough'', and what observables should
one look at?

\section{\large Critical Behavior: Fluctuations and Correlations}

Along the line of the second-order transition and at the tricritical point, 
thermodynamic observables exhibit continuous critical behavior; this implies 
in particular that higher order derivatives of the thermodynamic potential 
diverge in a well-defined functional form specified in terms of the critical 
exponents of the relevant symmetry group. These derivatives, in turn,
express calculable fluctuations of physical observables. For the idealized 
case considered here, the chiral limit of QCD, we thus obtain predictions 
for the fluctuations of baryon number, charge and strangeness. 

\medskip

\begin{figure}[htb]
\centerline{\epsfig{file=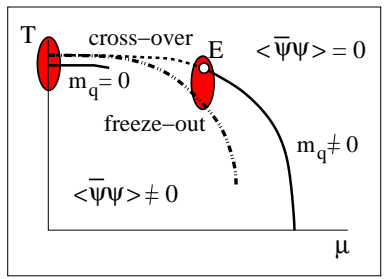,width=5cm}}
\vspace*{0.2cm}
\caption{Phase space regions proposed to look for remnant critical 
behavior.}
\label{remnant}
\end{figure}

There are two regions of critical behavior of particular interest for 
heavy ion studies. If the pseudocritical cross-over line for physical values 
of $m_q$ and small $\mu$ is sufficiently close to that of the second order 
transition in the chiral limit, we may there expect remnants of O(4) 
criticality, and we can look for them at top RHIC 
energy and at the LHC. At larger $\mu$ and finite $m_q$, we expect to  
encounter critical behavior near the end point $E$ of the first order 
regime, and the main aim of present beam energy scans at RHIC as well 
as that of the NA61 experiment at the CERN-SPS is the search for this end 
point through its effect on the freeze-out line. Here the exponents are 
those of the $Z_2$ group. In both cases, near $\mu=0$ and near the 
critical endpoint at finite $\mu$, we are thus looking for {\sl remnants of
criticality} (see Fig.\ \ref{remnant}). The pseudo-critical line near 
$\mu=0$ approaches the 
critical line only in the chiral limit $m_q \to 0$, and the freeze-out
line at finite $\mu$ is off the line of first order transition and the
critical endpoint by the presence of non-resonant interactions. 

\medskip

\begin{figure}[htb]
\centerline{\epsfig{file=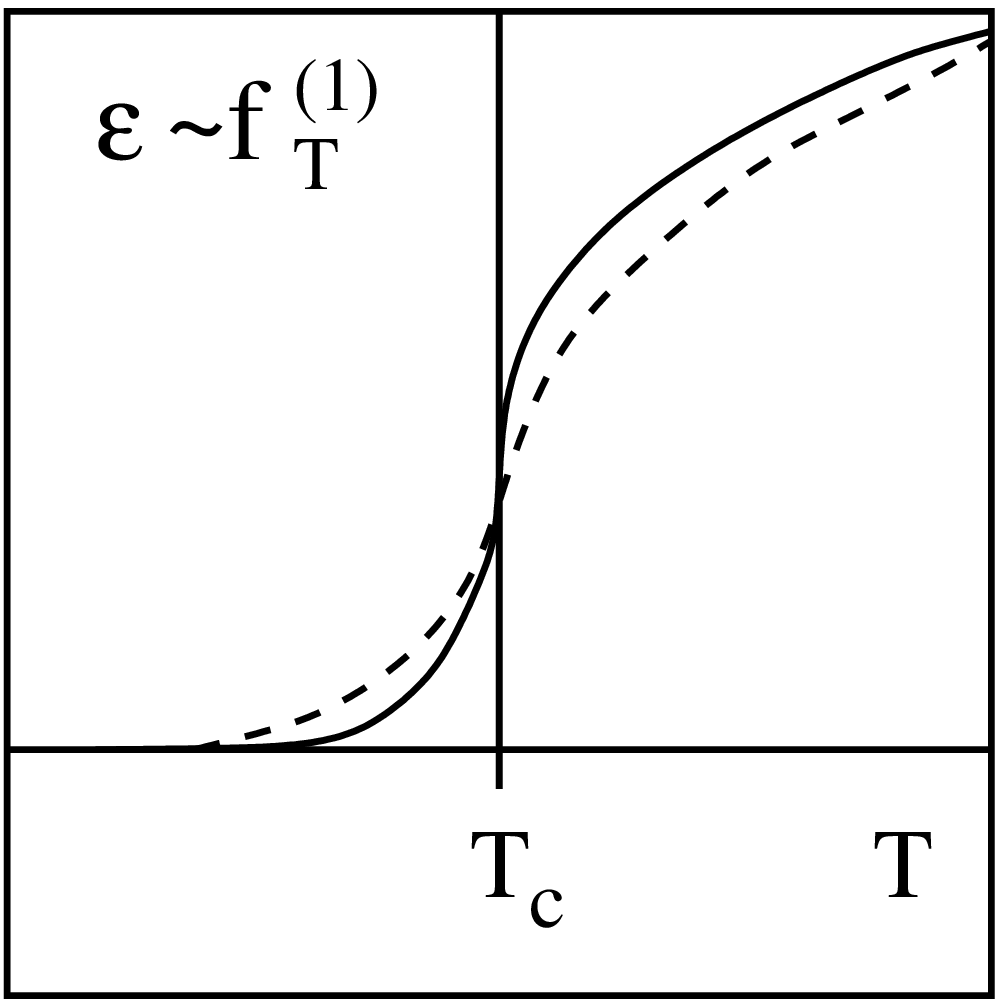,width=3.3cm}\hskip0.6cm
\epsfig{file=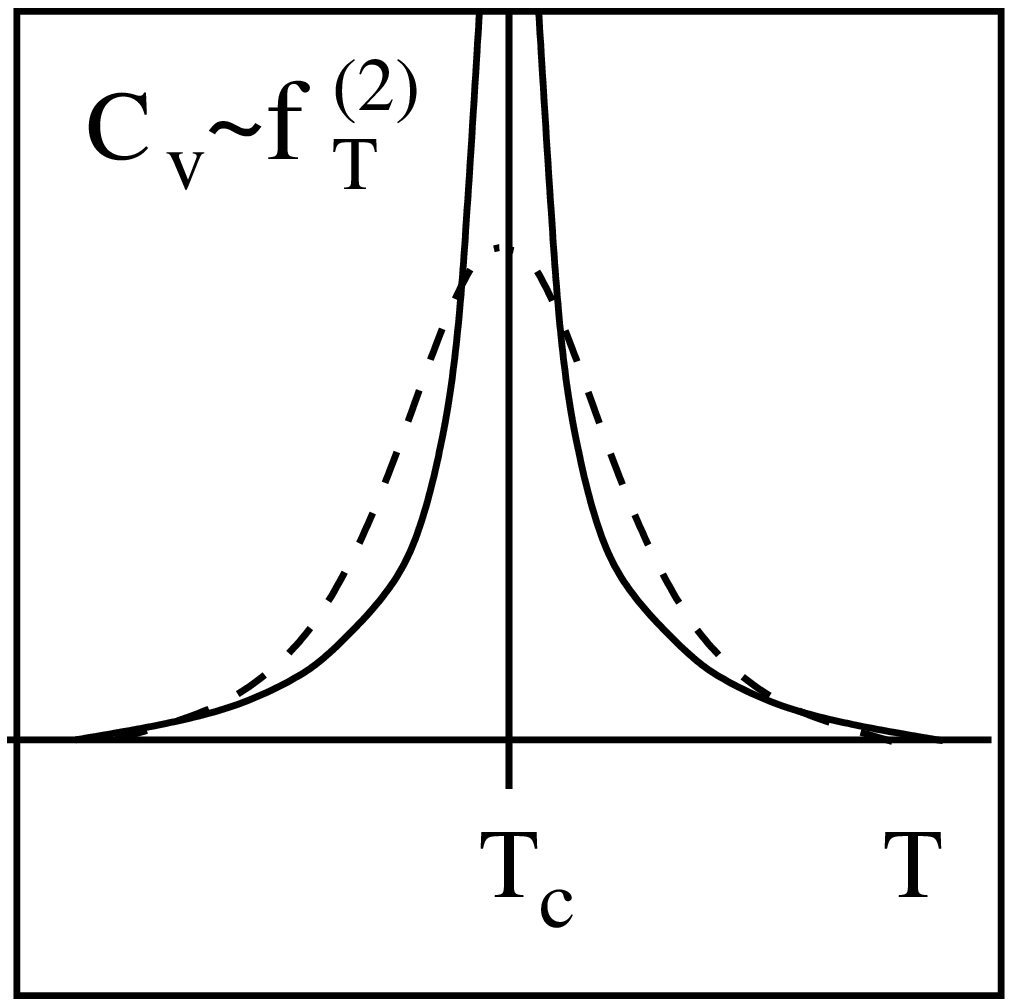,width=3.3cm}\hskip0.6cm
\epsfig{file=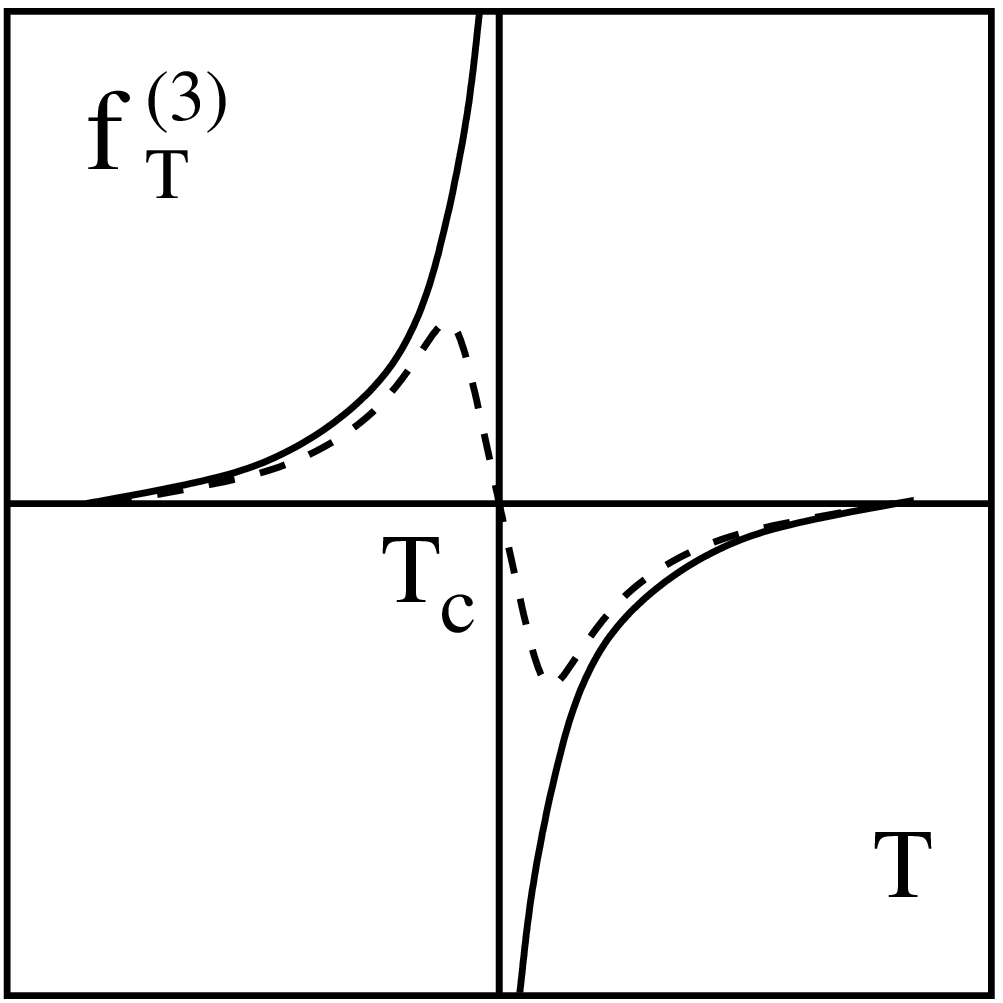,width=3.3cm}\hskip0.6cm
\epsfig{file=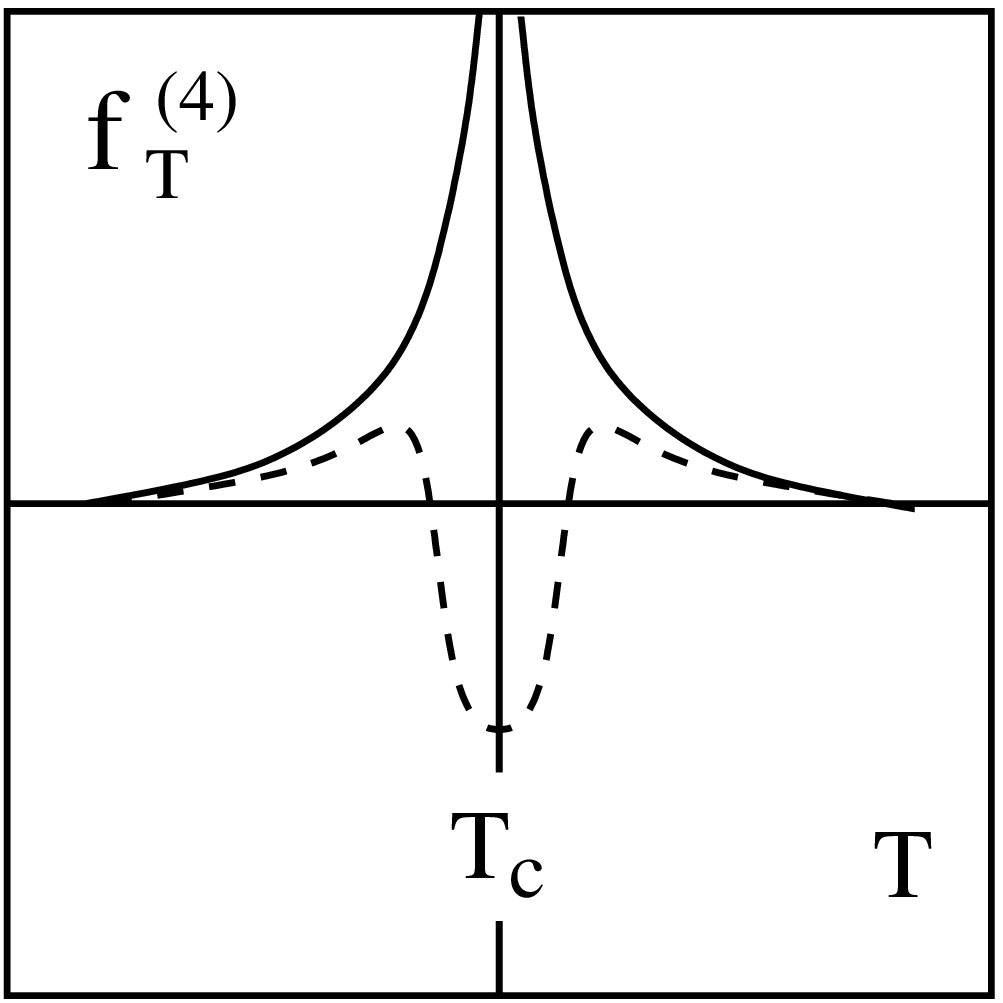,width=3.3cm}}

\vspace*{0.2cm}

\hskip1.9cm (a) \hskip3.3cm (b)\hskip3.3cm (c)\hskip3.4cm (d)

\caption{3d Ising model response functions up to fourth order,
for $H=0$, solid line, and for small finite external field $H$, dashed line.}
\label{crit}
\end{figure}

What can we calculate and what can we look for? Since the overall critical
structure of QCD as function of $T,~\mu$ and $m_q$ is somewhat involved,
let us first look at a simpler case for illustration, the Ising model of spins 
$s_i=\pm 1$ on a three-dimensional spatial lattice of $N^3$ sites.  It is 
defined by the Hamiltonian 
\be
{\cal H}~=~-J \sum_{\{i,j\}}s_is_j - H\sum_i s_i,
\label{2.1}
\ee
where $J$ specifies the interaction of next-neighbor spins and $H$ a possible
external field. For $H=0$, the system is $Z_2$ symmetric; the Hamiltonian
then remains invariant under spin flip $s_i \to -s_i~\forall~i$.
 The thermodynamics is encoded in the partition function   
\be
Z(T,H) = \prod_{i=1}^{N^3} \sum_{s_i=\pm 1} \exp{\{-\beta \cal H\}},
\label{2.2}
\ee
where $\beta=1/T$ is the inverse temperature; this in turn leads to the 
density of the (Helmholtz) free energy
\be
f(T,H) =  -{T\over V} \log Z(T,H),
\label{2.3}
\ee
with $V=N^3$. At first sight, it seems to vary smoothly with $T$ and $H$; 
to show explicitly the critical behavior inherent in the system, we have 
to look at derivatives with respect to $T$ and $H$, at the so-called 
{\sl response functions}. The first temperature derivative specifies the 
energy density
\be
\e(T,H=0) \sim \left({\partial f(T,H) \over \partial T} \right)_{H=0} 
\label{2.4}
\ee
and still shows continuous behavior, but the second, the specific heat,
\be
c_v (T,H=0) \sim \left({\partial \e(T,H) \over \partial T} \right)_{H=0} 
\sim \left({ \partial^{~\!2} f(T,H) \over \partial T^2} 
\right)_{H=0} \sim ~|T-T_c|^{-\alpha} \sim |t|^{-\alpha},  
\label{2.5}
\ee
diverges as a power at a critical temperature $T_c$; hence we use from now 
on $t=(T-T_c)/T_c$ as suitable variable, with the critical exponent $\alpha$ 
specifying the singular behavior. In Fig.\ \ref{crit}, we illustrate the
resulting patterns schematically. Included are here also the next two
derivative orders, and in all cases, we also show the expected pattern
if the symmetry is slightly broken by the presence of a small external 
field $H$. We note in particular the non-monotonic behavior of all 
derivatives higher than the energy density; this behavior signals
criticality and clearly deviates from any resonance gas pattern.
 
\medskip

The other variable $H$ leads to a similar pattern. The first derivative
with respect to $H$ gives at $H=0$ the spontaneous magnetisation,
\be
m(t,H=0) \sim \left( {\partial f(T,H) \over \partial H}\right)_{H=0}
\sim ~|t|^{~\!\beta},
\label{2.6}
\ee
for $t\!<\!0$; for $t\!\geq\!0$, $m(t,H=0)=0$: the magnetisation is the
order parameter and thus vanishes above the critical point for $H=0$.
For finite $H$, it remains finite even there, vanishing as
\be
m(t=0,H) \sim \left( {\partial f(T,H) \over \partial H}\right)_{t=0}
\sim h^{1/\delta},
\label{2.7}
\ee
along the critical isotherm, with $h=H/T_c$. The second derivative
\be
\chi_T(t,h=0) \sim \left( {\partial m(t,h) \over \partial h}\right)_{h=0}
\sim \left( {\partial^2 f(t,h) \over \partial h^2}\right)_{h=0}
\sim |t|^{-\gamma},
\label{2.7a}
\ee
gives the isothermal susceptibility, the rate at which the magnetisation
vanishes at the Curie point; it also diverges there.
In summary: critical behavior 
means that higher order derivatives of the pressure diverge in a 
functional form specified by critical exponents (we here had $\alpha,
~\beta,~\gamma$ and $\delta$) which are fixed once the symmetry group 
is given.

\medskip

The singular behavior of the response functions can in turn be related to
that of fluctuations and correlations. The specific heat determines
the energy fluctuation over the lattice,
\be
c_V(T) \sim \langle ( \sum_{i,j}s_is_j)^2 \rangle - 
\langle \sum_{i,j}s_is_j \rangle^2
\label{2.8}
\ee
while the susceptibility measures the fluctuation of the spin,
\be
\chi_T(t,H=0) \sim \langle ( \sum s_i)^2 \rangle - \langle \sum s_i \rangle^2;
\label{2.9}
\ee
in the absence of spin-spin correlations, it vanishes. The divergence of
the response functions at the critical point is thus connected to 
fluctuations diverging there, which in turn is a consequence of diverging 
correlations: at the critical point, the spin-spin correlation length
$\xi$ diverges as
\be
\xi(t) \sim |t|^{-\nu},
\label{corre}
\ee
so that constituents of all scales become correlated (critical opalescence).
Given sets of 
lattice configurations at various temperatures, we can thus, through the 
calculation of response functions, determine the onset and the analytical 
form of the critical behavior shown by the system. Using the form
(\ref{corre}),
one can also express the response functions in terms of the correlation
length. For the 3d Ising model, the specific heat becomes
\be
\chi_T \sim \xi^{\gamma/\nu} \sim \xi^2,
\label{corre2}
\ee
using $\gamma \simeq 1.2,~\nu \simeq 0.6$ for the 3d Ising model;
higher derivatives grow as higher powers of $\xi$.

\medskip

In the case of QCD, we have in addition to the temperature the chemical 
potentials of the conserved quantum numbers as thermodynamic parameters,
$\mu_B$ for the baryon number, $\mu_Q$ for the electric charge, and $\mu_S$
for the strangeness, respectively:
\be
f_{\rm Ising}(T,H)~ \to~P_{\rm QCD}(T, \mu_B, \mu_Q, \mu_S,m_q).
\label{2.10}
\ee
The one-dimensional temperature space $T$ is thus generalized to a 
four-dimensional space $T,\mu_B,\mu_Q,\mu_S$; variations of the 
chemical potentials do not affect the intrinsic symmetry of the system.
The light quark mass $m_q$ plays, as
mentioned, the role of the external field: for $m_q\not= 0$, the chiral
symmetry of the Lagrangian is broken explicitly. To illustrate the effect
of additional chemical potentials, we consider the baryon number
case, simplifying the notation, $\mu_B=\mu$. The critical point in $T$  
thus now becomes a critical line in the $T-\mu$ plane, which renders the
relation between variables and critical exponents somewhat more complex,
see e.g. \cite{t-mu}. The principle remains the same, however: to find
experimental evidence of criticality, we have to look for non-monotonic 
behavior of response functions measured in heavy ion collisions, to be
compared to such behavior obtained in lattice studies. The finite interaction 
volume in actual collisions, together with time evolution effects, will
limit the maximum size of correlations and thus mask a possible divergence.
Since the higher derivatives depend on higher powers of $\xi$, they could
provide a more sensitive tool.

\medskip

\begin{figure}[htb]
\centerline{\epsfig{file=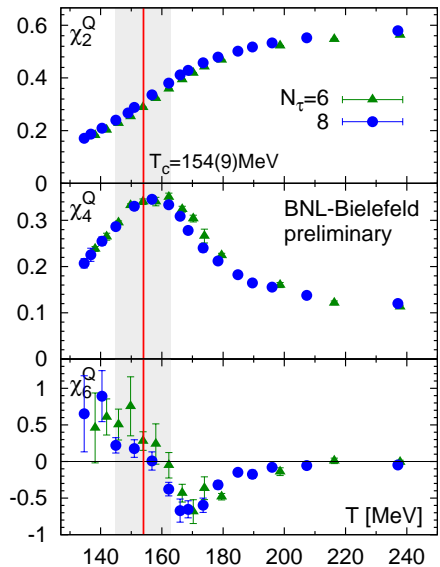,width=6cm}}

\vspace*{0.2cm}


\caption{Schematic behavior of the fourth ($\chi_4^Q$) and sixth ($\chi_6^Q$) 
order cumulants of the electric charge in the critical region, for 
resonance gas (solid line) vs. lattice QCD calculations (dashed line) 
\cite{CS}.}

\label{charge}
\end{figure}

The first task is thus to determine in lattice QCD some evidence for
critical behavior in the hadronic state, behavior deviating from that of
an ideal resonance gas. Here it has to be noted that the partition
function in QCD depends on {\sl squared} electric charges and baryochemical
potential, since it is left invariant under a change of sign of these
quantities. As a result, the non-monotonic behavior shown for the Ising
model is shifted to higher order cumulants: the third order spin behavior 
is expected for the sixth order in QCD, and so on. 
So far, derivatives up to the second order were 
seen to agree with the non-critical resonance gas, see Fig.\ \ref{HRG}(a).
First evidence for deviations has been found quite recently \cite{CS},
studying the sixth order cumulant of the electric charge ($\chi^Q_6$); 
the result is illustrated schematically in Fig.\ \ref{charge}. We see that 
at the critical point, i.e., at the freeze-out value of the resonance gas, 
the sixth order cumulant vanishes for QCD, in contrast
to the continued monotonic increase expected from the hadronic resonance gas. 
Such differences are expected to continue for higher orders: the eighth 
cumulant should become negative in critical QCD, large and positive for the 
resonance gas. The onset of critical behavior, difficult if not impossible 
to detect in lower order cumulants, should thus become more and more evident 
with increasing order.

\medskip

We should here emphasize that the forms shown in Fig.\ \ref{charge} are
obtained in physical finite temperature lattice studies, i.e., for 2+1
quark flavors of physical masses, and they give the behavior of the full
cumulants, not just a singular part. They are therefore {\sl bona fide}
predictions: if these quantities become measurable in heavy ion collisions
and there do not show the predicted form, the produced systems are not 
governed by equilibrium QCD thermodynamics. There can, of course, be 
various reasons for why this might be the case, up to Feynman's broken
watch, but the result as such would be an observation decisive for our 
understanding of high energy nuclear collisions.

\medskip

The search for the critical endpoint will require a similar program, here
focused on the baryon density, to be varied by varying the collision
energy. The data obtained will again be analysed in terms of the ideal 
resonance gas, even though we now expect some non-resonant baryon-baryon 
interactions. And any onset of criticality should produce deviations from 
monotonic behavior under variations of the collision energy.

\medskip

\begin{figure}[htb]
\centerline{\epsfig{file=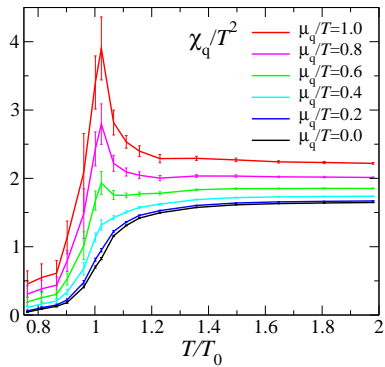,width=5cm}}
\vspace*{0.2cm}
\caption{Baryon number susceptibility in two-flavor QCD \cite{Ejiri}.}
\label{barsus}
\end{figure}

Predictions from lattice QCD here are not as readily obtained, since
the conventional simulation scheme breaks down for finite $\mu$. Several
extensions have been proposed to calculate at least up to some small but
finite values. A power series expansion in the baryochemical potential,
retaining terms up to second order, gives for cumulant governing 
baryon number fluctuations the form
\be
\chi_B^{(2)}(t,\mu)  \sim P^{(2)}(t,\mu=0) + \mu^2 P^{(3)}(t,\mu=0),
\label{chiB}
\ee  
with $t=(T-T_c)/T_c$. The first term is essentially the energy density,
the second contains the specific heat. In the case of $Z_2$
critical behavior, the latter diverges at $t=0$, and so lattice studies
find the pattern shown in Fig.\ \ref{barsus}, indicating an onset of
non-monotonic variation around the critical temperature. The use of the 
Taylor expansion in $\mu$ does not allow this scheme for a determination
of the critical point.
 
\medskip

The remnant critical behavior of QCD thermodynamics thus is encoded in
the higher cumulants of conserved quantum number distributions. Theoretically,
these have been and are being studied in considerable detail, for baryon
number, electric charge and strangeness. To what extent these are measurable
in heavy ion collisions is another issue, beyond the scope of this report.
The high statistics experiments at the LHC and the extended beam scan efforts 
at RHIC give rise to hope for precise enough data, e.g. of charge and baryon 
number distributions, to allow such studies.  

\section{\large The QGP Temperature: Quarkonium Suppression}

The ultimate task for nuclear collision studies is evidently to probe if the 
produced medium in its early pre-hadronic stage is indeed the quark-gluon
plasma of QCD. This requires probes present at the pre-confinement stage,
and three such probes have been proposed and discussed. Electromagnetic
radiation formed in quark interactions can escape from the medium unaffected 
and thus deliver information about the medium at the time of its formation.
Such radiation will, however, also be produced in the hadronic 
state, and an identification of its origin is difficult. Hard jets are 
formed at a ``hard'' early time and must pass through the subsequent 
pre-hadronic (and hadronic) medium to reach detectors. The energy loss 
of these jets will 
reflect the medium being passed, although there are so far no quantitative
QCD predictions for this. Heavy charm and bottom quark-antiquark pairs will 
also be formed at an early time of a scale set by the heavy quark mass. 
Their fate -- open charm/bottom or quarkonium -- will reflect the temperature 
of the medium, indicating if it is too hot for binding or not \cite{MS}. Here 
extensive studies have been carried out over the years, and these will
be my final topic.
  
\medskip

Because of the heavy mass of charm and bottom quarks, quarkonium spectra
can in good approximation be calculated in non-relativistic potential
theory; this reproduces the observed (spin-averaged) masses of the
ground states as well as of the different excited states to better than 
a few percent \cite{Schrodinger}. The resulting binding energies are quite 
large and the binding radii small, with 600 MeV and 0.2 fm for the 
charmonium ground state \J, 1.2 GeV and 0.1 fm for the bottomonium ground
state \U. Hence one expects that even in a deconfined 
medium just above the transition temperature (where the screening radius 
is around 1 fm), they can still survive as bound states. Only a considerably 
hotter QGP will eventually prevent binding, both through color screening 
and through collision dissociation. Raising the temperature of the initial 
medium through an increase of the collision energy will thus give rise to a 
step-wise suppression of quarkonia. First the more weakly bound, larger
excited states (1P, 2S, ...) will no longer be present, until eventually
even \J~and \U~must become suppressed.

\begin{figure}[htb]
\centerline{\epsfig{file=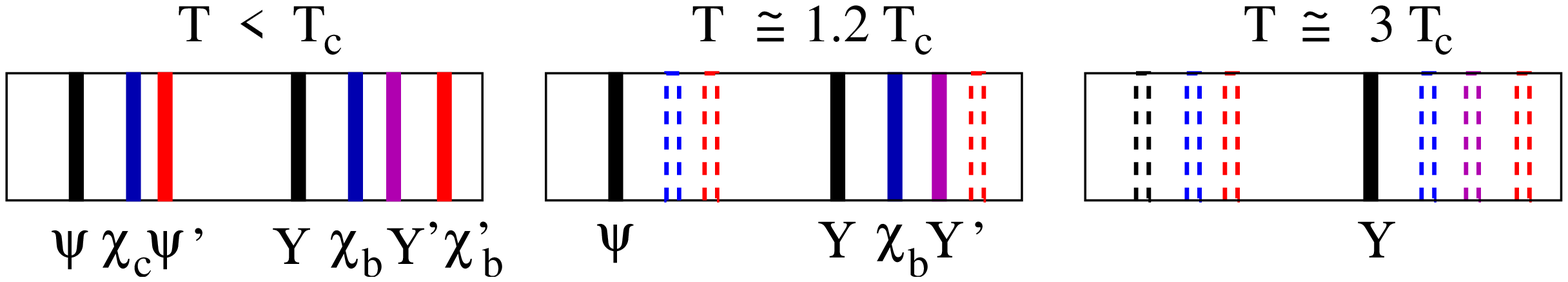,width=10cm}}
\caption{Quarkonium spectral lines as QGP thermometer} 
\label{spectral}
\end{figure}

To give an indication of the process, we consider the potential theory
approach, using the charm system for illustration. The Schr\"odinger equation
\be
\left\{2m_c -{1\over m_c}\nabla^2 + V(r)\right\} \Phi_i(r) = M_i \Phi_i(r),
\label{schroedinger}
\ee
with the ``Cornell'' form for the confining potential \cite{Cornell},
\be
V(r) = \sigma ~r - {\alpha \over r},
\ee 
in terms of charm quark mass $m_c$, string tension $\sigma \simeq 0.2$ 
GeV$^2$ and gauge coupling $\alpha \simeq \pi/12$, then determines the 
masses $M_i$ and the radii $r_i$ of the different charmonium states in
vacuum. In a hot QGP, the potential is replaced by the color-screened
form
\be
V(r,T) \sim \sigma r \left\{ {1-e^{-\mu r} \over \mu r} \right\} 
- {\alpha \over r} e^{-\mu r}, 
\label{schwinger}
\ee
where $\mu(T)$ specifies the screening mass of the medium and hence 
$r_c(T)=1/\mu(T)$ the screening radius. When this falls below the binding
radius of of a given charmonium state, $c$ and $\bar c$ no longer ``see'' 
each other and a binding for that state is not possible. Approximating 
$\mu(T)$ by the form obtained in heavy quark lattice studies leads to 
suppression thresholds
\be
T_{\j}~\simeq~1.3~T_c,~~~ T_{\x}~\&~T_{\p}~\simeq~1.1~T_c.
\ee
The thresholds for the bottomonium states are shifted to 
correspondingly higher tempe\-ra\-tures. 

\medskip

Such a potential theory treatment is, of course, quite phenomenological.
Many attempts have replaced the model input potential (\ref{schwinger})
by a form obtained directly from heavy quark lattice studies. This still 
retains ambiguities, however, and it does not include the effect of 
collision dissociation, which results in an imaginary part of the potential. 
The {\sl ab initio} approach is to calculate the in-medium quarkonium 
spectrum directly in finite temperature lattice QCD, and since several years 
that endeavor is under way by various groups 
\cite{Umeda,Asakawa,Datta,Iida,Jacovac,Skullerud,Ding}
In these calculations, the 
quarkonium correlator $G(\tau,T)$ is determined at temperature $T$; it is 
an integral transform of the desired quarkonium spectrum $\sigma(\omega,T)$,
\be
G_i(\tau,T) = \int d\omega~ \sigma_i(\omega,T)~ 
{\cosh[\omega(\tau - (1/2T))] \over \sinh(\omega/2T)}.
\label{lspec1}
\ee
In principle, the transform just has to be inverted to obtain the spectrum;
in practice, the correlator is given only on for a finite number of points
in $\tau$, which makes the inversion ambiguous. It is therefore generally
carried out by the {\sl Maximum Entropy Method}, a scheme to reconstruct
something based on fragmentary information \cite{MEM}.
So also here the last word is 
not yet said, though with increased computer performance and size, the 
results are constantly gaining in certainty. Some as yet not final, though
perhaps indicative results are shown in the following table.

\medskip

\begin{center}
\renewcommand{\arraystretch}{1.5}
\begin{tabular}{|c||c|c|c||c|c|c|c|c|}
\hline
 state & J/$\psi(1S)$ & $\chi_c$(1P) & $\psi(2S)$&$\Upsilon(1S)$&
$\chi_b(1P)$&$\Upsilon(2S)$&$\chi_b(2P)$&$\Upsilon(3S)$\\
\hline
\hline
$T_d/T_c$ & 1.5  & 1.1 & 1.1 & $>4.0$ & 1.8& 1.60& 1.2 & 1.1 \\
\hline
\end{tabular}\end{center}

\medskip

\begin{figure}[htb]
\centerline{\epsfig{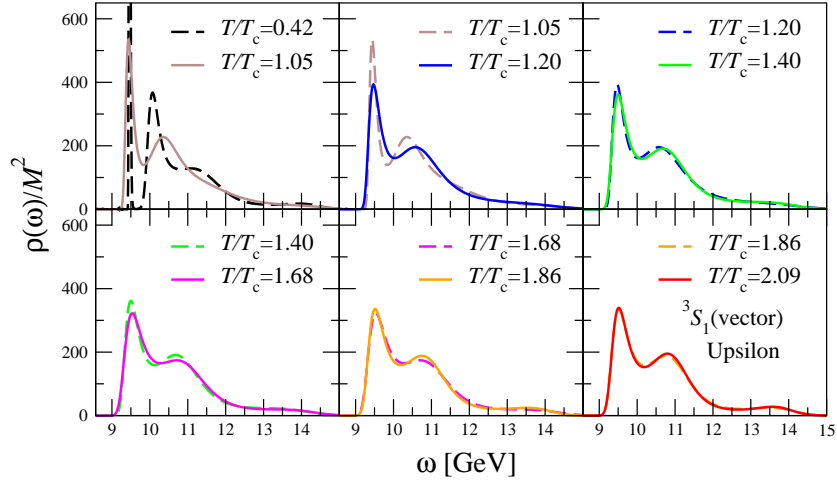}}
\caption{Suppression patterns for bottomonia in NRQCD \cite{Aarts}}
\label{exbot}
\end{figure}

For bottomonium, a recent temperature scan \cite{Aarts} based on 
lattice studies in non-relativistic QCD provides some more details
of the suppression pattern for the higher excited states. The results
are shown in Fig.\ \ref{exbot}. 

\medskip

Once the final answers are given, we can specify up to what
temperature the survival of a given quarkonium state in a hot QGP is
possible, and moreover give the ratios of the different survival thresholds.

\medskip

\begin{figure}[htb]
\centerline{\epsfig{file=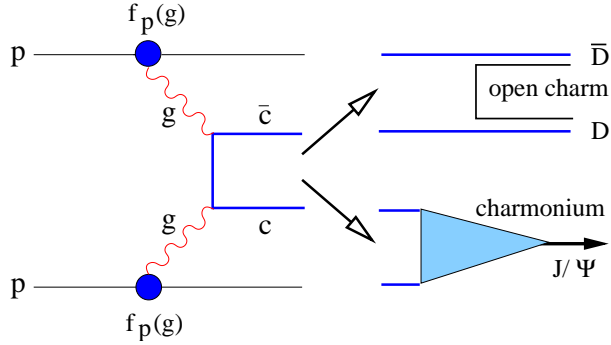,width=8cm}}
\caption{Schematic view of \J~production in $pp$ collisions}
\label{pp}
\end{figure}

To discuss the effect of this on quarkonium production, we first recall 
underlying dynamics, again using the \J~for illustration.
The production process in elementary
hadronic collisions (taking $pp$ as example) begins
with the formation of a $\C$ pair; this pair can then either lead to
open charm production (about 90 \%) or subsequently bind to form
a charmonium state (about 10 \% for all charmonia). A schematic 
illustration (Fig. \ref{pp}) shows the dominant high energy reaction 
through gluon fusion. 

\medskip

The initial $\C$ production can be calculated in terms of the parton 
distribution functions $f_p$ of the relevant hadrons and the pertubative
partonic cross section. The full description of charmonium binding has 
so far resisted various theoretical attempts; on the other hand, the
process is in good approximation independent of the incident hadronic
collision energy \cite{Gavai,HP}. This is a consequence of the fact that 
the heavy quark propagator in the reaction $gg \to \C$ strongly dampens 
the mass variation of the $\C$ pair with incident energy. Thus the
fractions of the produced $\C$ system into hidden vs.\ open charm
as well as those for the different charmonium states are approximately
constant; once determined at one energy, they remain the same also for
different collision energies. As a result, the phenomenological color 
evaporation model \cite{CE1,CE2,CE3,CE4} provides a good description 
of charmonium production through the form
\be
\sigma_{hh\to \j}(s) = g_{\C \to \j}~\!\sigma_{hh \to \C}(s),
\label{cem}
\ee
and correspondingly for the other charmonium states. Here
the constant $g_{\C \to \j}$ specifies what fraction of the total
$\C$ production cross section goes into \J~production; in $pp$
collisions it is typically about 2 \%. The set of the different
constants $g_{\C \to i}$ for the different charmonium states $i$
thus effectively characterizes charmonium production in the absence
of a medium. 

\medskip

A further important aspect of quarkonium production in elementary
collisions is that the observed (1S) ground states \J~and \U~are in 
both cases partially produced through feed-down from higher excited 
states \cite{FD1,FD2,FD3,FD4}. Of the observed \J~rates, only some
60\ \% is a directly produced $\j(1S)$ state; about 30 \% comes from 
$\chi_c(1P)$ and 10 \% from $\psi'(2S)$ decay. Because of the narrow 
width of the excited states, their decay occurs well outside any 
interaction region.

\medskip

The features we have here summarized for charm and charmonium production
are readily extended to that of bottom and bottomonium. To simplify
the discussion, we shall continue referring to the charmonium case,
keeping in mind that all arguments apply as well to bottomonia. Given 
the patterns observed in elementary collisions, we want to see how they 
are modified in the presence of a medium, as provided by nuclear collisions. 
From the point of view of production dynamics, one way such modifications 
can arise is as {\sl initial state effects}, which take place before
the $\C$ pair is produced. The main possibilities considered so far are
nuclear modifications of the parton distribution functions (shadowing or
antishadowing) and a possible energy loss of the partons passing through
the nuclear medium to produce the $\C$. Once produced, the pair can encounter 
{\sl final state effects}, either in the form of a phase space shift 
already of the $\C$, e.g., through an energy loss of the unbound
charm quarks, or through effects on the nascent or fully formed
charmonium state.  Such effects may arise from the passage through the
cold nuclear medium, or because of the presence of the medium newly
produced in the nuclear collision. The latter is evidently what we have
in mind when we want to use quarkonia to study quark-gluon plasma
production. 

\medskip

\begin{figure}[htb]
\vspace*{-0mm}
\centerline{\psfig{file=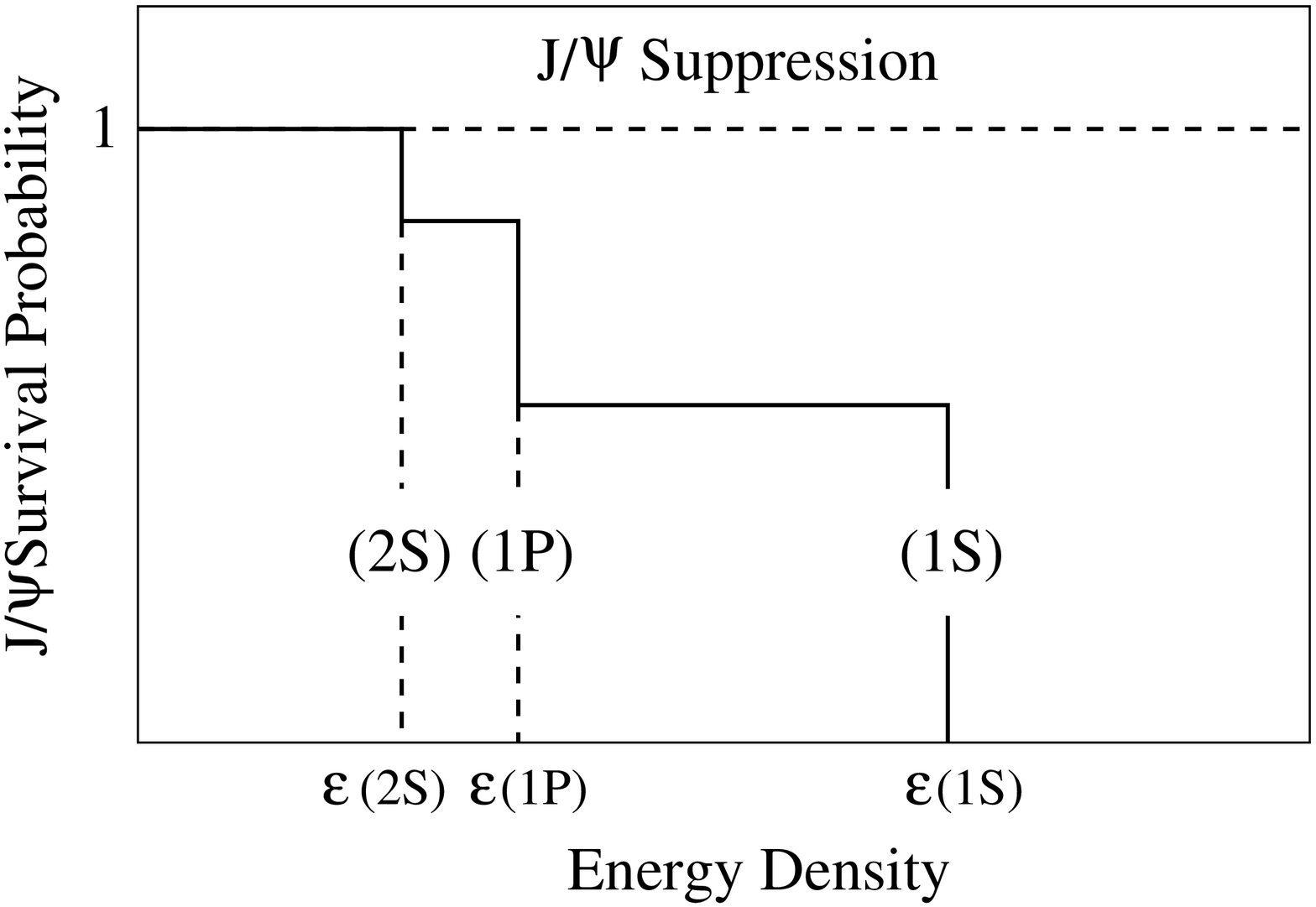,width=6cm,height=5cm}\hskip1cm
\psfig{file=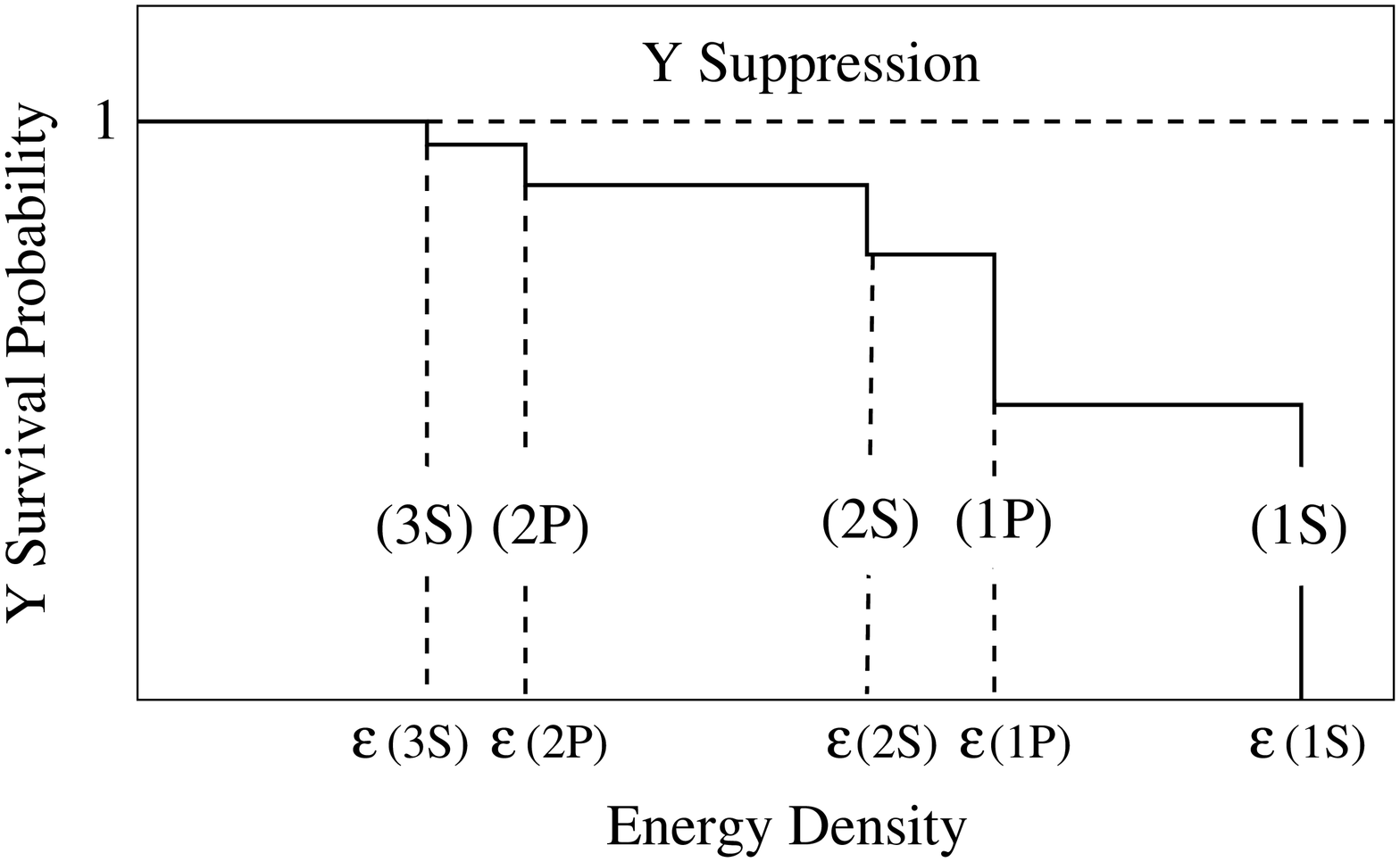,width=6cm,height=5cm}~~~}
\caption{Sequential quarkonium suppression}
\label{seq}
\end{figure}

Next we turn to quarkonium binding in a hot medium.
Color screening in a quark-gluon plasma will decrease the binding
force, both in strength and in its spatial range, and this 
should for sufficiently energetic nucleus-nucleus collisions suppress
quarkonium formation. In addition, there will be collision break-up.
Since the larger and less tightly bound states will be suppressed
at lower temperature or energy density than the ground states, the result
will be {\sl sequential suppression} \cite{Seq1,Seq2}. We illustrate this
for the \J. After an initial threshold suppressing the \P~and hence remo\-ving
its feed-down component for \J~production, there will be a second threshold
for $\chi_c$ suppression and then finally a third, at which the direct 
$\j(1s)$ is dissociated. The resulting pattern is illustrated in Fig.\ 
\ref{seq}. We have here introduced something denoted as \J~survival 
probability: the chance of a \J~to persist as a bound state in a deconfined 
medium. A similar sequential suppresssion pattern will arise for 
the step-wise removal for the higher state contributions to \U~production;
both quarkonium patterns are shown in Fig.\ \ref{seq}.

\medskip
 
We had already introduced the concept of quarkonium survival; that has
to be specified more explicitly.
Since we are interested in using quarkonium production as a tool to
study the medium produced in nuclear collisions, our primary concern
is not if such collisions produce more or fewer $\C$ pairs than 
proton-proton collisions, but rather if the presence of
the medium modifies the fraction of produced $\C$ pairs going into
charmonium formation. In other words, the crucial quantitity is the 
amount of charmonium production relative to that of open charm
\cite{Sridhar,HS13}. 
To illustrate: in $pp$ collisions, about 2 \% of the total $\C$ production
goes into \J's. If in high energy nuclear collisions the total $\C$
production rate were reduced by a factor two, but we still have 2 \%
of these going into \J's, then evidently $AA$ collisions do not modify
\J~binding. Hence the relevant
observable is the fraction of charmonia to open charm, or more generally,
that of quarkonia to the relevant open heavy flavor production 
\cite{Sridhar,PBM1,HS13}. In this quantity, if measured over the entire phase 
space, the effects of possible initial state nuclear modifications -- 
shadowing/antishadowing, parton energy loss -- cancel out, so that
whatever changes it shows relative to the $pp$ pattern is due to final
state effects.  

\medskip

Possible alternative variables to consider are the production ratios of 
the different quarkonium states; this has very recently become of particular
interest for bottomonium studies. Since \U(1S), \U(2S) and \U(3S) 
lie quite close to each other in mass, in the ratio of their rates again
initial state effects are expected cancel out.

\medskip

Let me close by commenting briefly on the experimental status. \J~suppression
in nuclear collisions, when compared to $pp$ collisions, was observed from
the very beginning. Those data have, however, remained inconclusive for 
some twenty-five years, essentially because the crucial observables were
not available. Lack of open charm data was compensated by comparison to
$pp$ data, and this brought in uncontrollable uncertainties due to initial
state effects. These effects were measured in $pA$ interactions, modelled 
and then used to construct $AA$ predictions, which necessarily remained
model-dependent. Since a few years, open charm data have become available,
so that it is only a matter of time now before the crucial variables
can be experimentally determined.

\medskip

For charmonia, measuring open to hidden flavor is moreover of interest
for yet another reason.  The abundance of $\C$ pair production at LHC
energy has led to the suggestion that at hadronisation a $c$ from one
collision may statistically combine with a $\bar c$ from another, thus
providing a new secondary charmonium production mechanism. Even with
all primary \J~suppressed, such statistical combination could lead to
an abundant new production at hadronisation \cite{SR1,SR2,SR3,SR4}. 
Again here the prediction
is a reshuffling of the $\C$ pairs between hidden and open charm channels,
so that only a measurement hidden/open can provide an unambiguous answer.
If such a mechanism is indeed effective, the sequential pattern we
had discussed is no longer present; instead, the ratios are those
of the hadrosynthesis of the charm sector. Evidently, this would 
constitute strong evidence for a thermal quark-gluon plasma.

\medskip  

\begin{figure}[htb]
\centerline{\psfig{file=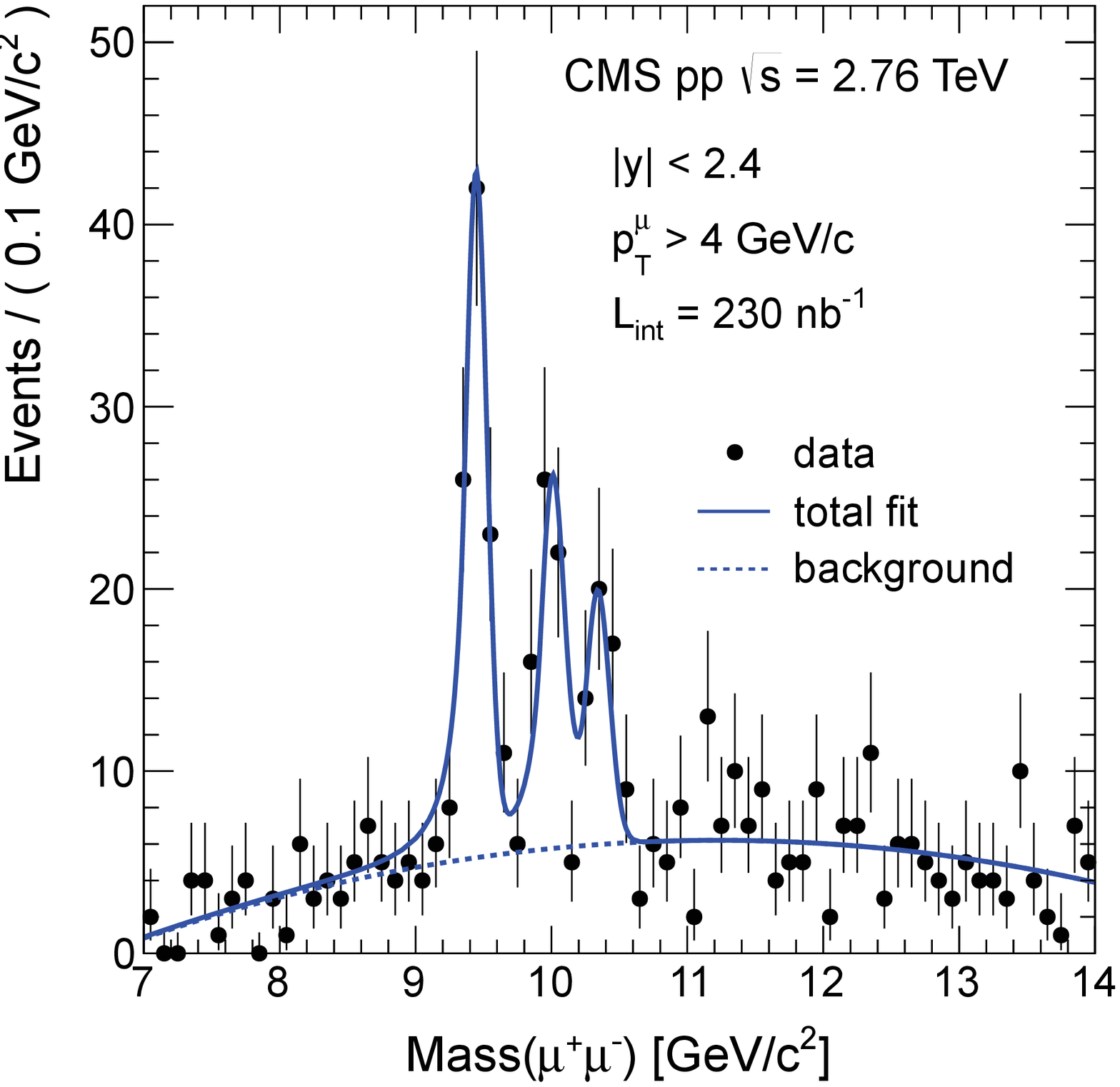,width=5cm}\hskip0.2cm
\psfig{file=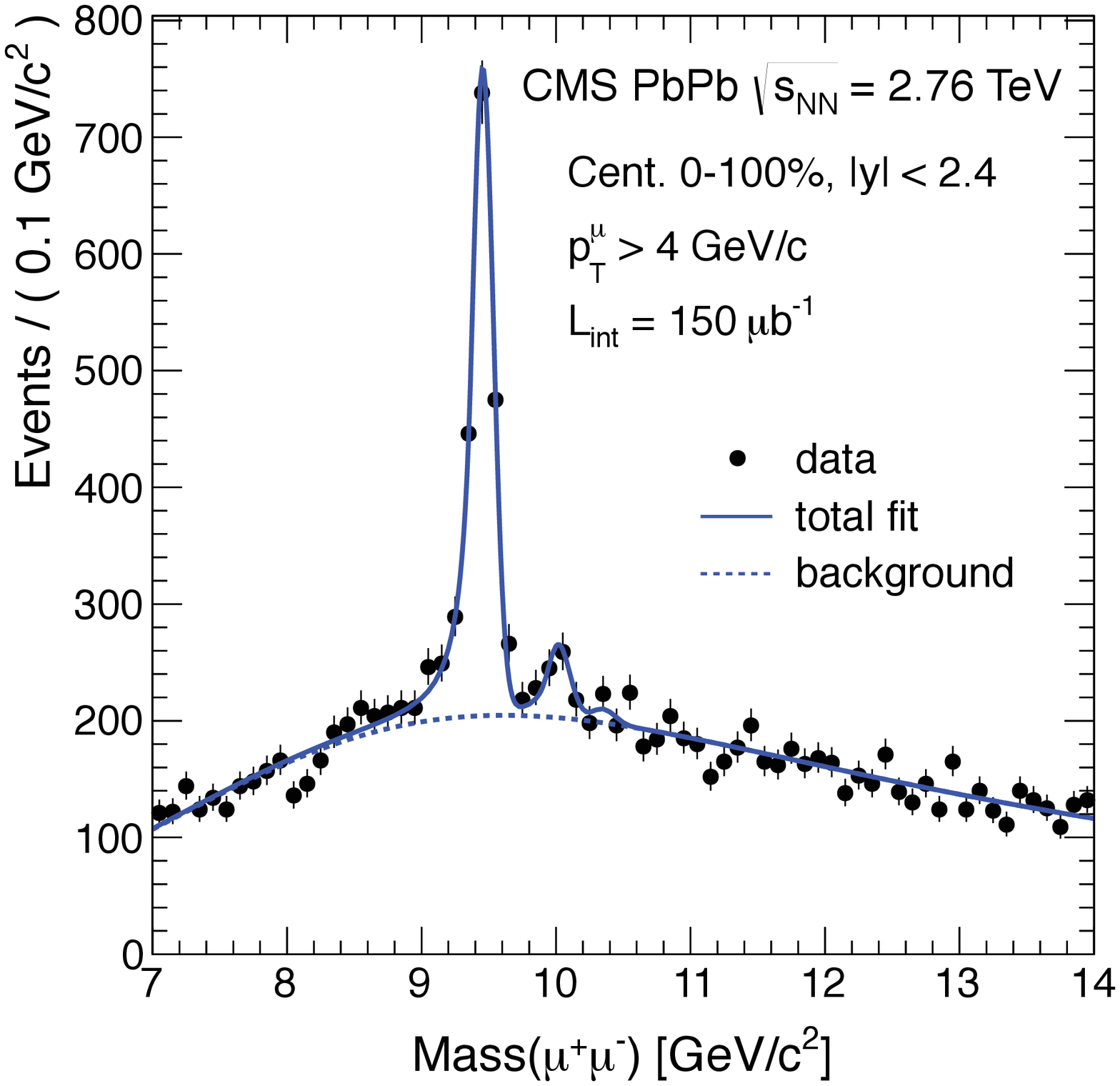,width=5cm}
\hskip0.2cm
\psfig{file=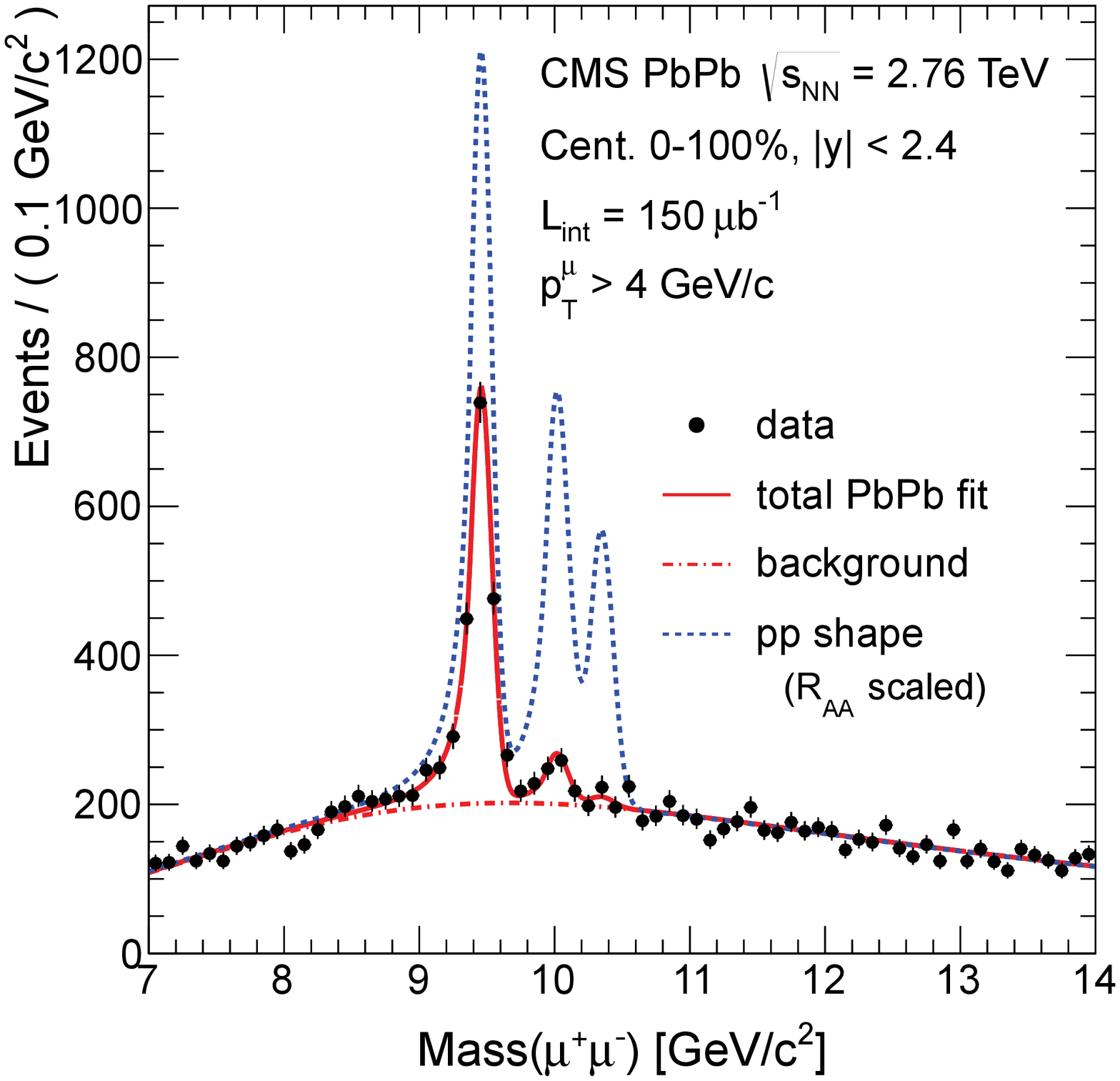,width=5cm}}
\caption{Bottomonium production in $pp$ (left) and $Pb-Pb$ (right) collisions,
as measured by the CMS collaboration at CERN--LHC \cite{CMS-u}}
\label{upseq}
\end{figure}

The role of the QGP thermometer would in that case be played by 
bottomonium production. We therefore close this section with some
recent data on this, data which provides perhaps the first clear
evidence for sequential suppression \cite{CMS-u}. In Fig.\ \ref{upseq} the
bottomonium spectrum in $pp$ collisions is compared to that in 
$Pb-Pb$ collisions, both measured at the LHC for $\sqrt s=2.75$ TeV.
In nuclear collisions, the higher excited states \U(2S) and \U(3S) 
are strongly suppressed relative to the ground state \U(1S). The 
production rate for this, however, is also found to be reduced there,
essentially by the contributions it received in $pp$ interactions
through feed-down from excited states. This becomes more evident 
in the right panel of
Fig.\ \ref{upseq}(c), where the scaled background curves are matched. 

\section{Conclusions}

Our aim was to focus on how in high energy nuclear collisions one can
test predictions from equilibrium statistical QCD. We showed in particular 
that
\begin{itemize}
\item{nuclear collisions at low baryon density produce a hadronic medium
in thermal equilibrium at the confinement temperature found in lattice QCD;}
\item{the critical behavior at the hadronization transition is encoded in 
fluctuations calculated in QCD, and these are in principle measurable
for baryon number, charge and strangeness;
\item{the suppression thresholds of quarkonium states specify the temperature
of the QGP; they are calculable in lattice QCD and measurable for 
charmonia (caveat: regeneration through statistical combination) and 
bottomonia.}
}
\end{itemize}

\end{document}